\documentclass[sigplan,10pt,preprint]{acmart}

\usepackage{algorithm}
\usepackage{algpseudocode}
\usepackage{listings}
\usepackage{multicol}
\usepackage{subcaption}
\usepackage{paralist}
\usepackage{bussproofs}
\usepackage{makecell}
\usepackage{pgfplotstable}
\usetikzlibrary{patterns}
\usepgfplotslibrary{statistics}
\usepackage{multirow}
\usepackage{titlesec}
\usepackage[skip=2pt]{caption}
\usepackage{tikz}
\usetikzlibrary{arrows.meta}
\colorlet{lightgray2}{gray!30}
\colorlet{lightgray1}{lightgray2!10}
\colorlet{darkred}{red!80!black}

\AtBeginDocument{%
  }

\setcopyright{acmcopyright}
\copyrightyear{2023}
\acmYear{2023}
\acmDOI{XXXXXXX.XXXXXXX}

\acmConference[CGO '23]{Make sure to enter the correct
  conference title from your rights confirmation emai}{Feb.25--Mar.03,
  2023}{Montreal, QC, Canada}
\acmPrice{15.00}
\acmISBN{978-1-4503-XXXX-X/xx/xx}

\newcommand{\coolname}{DFI}

\newcommand{\interval}[1]{\langle #1 \rangle}
\newcommand{\intervalset}[1]{\Pi_{#1}}

\newcommand{\numBenchmarks}{4}

\lstdefinestyle{MLIRStyle}{
  breaklines=true,
  basicstyle=\fontsize{7}{9}\selectfont\ttfamily,
  stepnumber=1,
  numbers=left,
  numbersep=6pt,
  xleftmargin=1em,
  escapechar=?,
  abovecaptionskip=0pt
}

\begin{document}

\titlespacing*\section{0pt}{3pt plus 1pt minus 1pt}{0pt plus 1pt minus 1pt}
\titlespacing*\subsection{0pt}{2pt plus 1pt minus 1pt}{0pt plus 1pt minus 1pt}
\titlespacing*\subsection{0pt}{2pt plus 1pt minus 1pt}{0pt plus 1pt minus 1pt}

\setlength{\abovedisplayskip}{2pt}
\setlength{\belowdisplayskip}{4pt}

\setlength{\floatsep}{3.0pt plus 1.0pt minus 3.0pt}
\setlength{\textfloatsep}{3.0pt plus 1.0pt minus 3.0pt}
\setlength{\intextsep}{3.0pt plus 0pt minus 3.0pt}

\title{\coolname{}: An Interprocedural Value-Flow Analysis Framework that Scales to Large Codebases}


\author{Min-Yih Hsu}
\affiliation{%
  \institution{University of California, Irvine}
  \country{United States}}
\email{minyihh@uci.edu}

\author{Felicitas Hetzelt}
\affiliation{%
  \institution{University of California, Irvine}
  \country{United States}}
\email{fhetzelt@uci.edu}

\author{Michael Franz}
\affiliation{%
  \institution{University of California, Irvine}
  \country{United States}}
\email{franz@uci.edu}


\begin{abstract}
  Context- and flow-sensitive value-flow information is an important building block for many static analysis tools.
  Unfortunately, current approaches to compute value-flows do not scale to large codebases, due to high memory and runtime requirements.
  This paper proposes a new scalable approach to compute value-flows
  via graph reachability.
  To this end, we develop a new graph structure as an extension of LLVM IR that contains two additional operations which significantly simplify the modeling of 
  pointer aliasing.
  Further, by processing nodes in the opposite direction of SSA def-use chains, we are able to minimize the tree width of the resulting graph.
  This allows us to employ efficient tree traversal algorithms in order to resolve graph reachability.

  We present a value-flow analysis framework, \coolname{}, implementing our approach. 
  We compare \coolname{} against two state-of-the-art value-flow analysis frameworks, Phasar and SVF,
  to extract value-flows from \numBenchmarks{} real-world software projects.
  Given 32GB of memory, Phasar and SVF are unable to complete analysis of larger projects such as \texttt{OpenSSL} or \texttt{FFmpeg}, while \coolname{} is able to complete all evaluations.
  For the subset of benchmarks that Phasar and SVF do handle,
  \coolname{} requires significantly less memory (1.5\% of Phasar's, 6.4\% of SVF's memory footprint on average) and
  runs significantly faster (23x speedup over Phasar, 57x compared to SVF).
  Our analysis shows that, in contrast to previous approaches, \coolname{}'s memory and runtime requirements scale almost linearly with the number of analyzed instructions.


\end{abstract}


\maketitle

\section{Introduction}

Value-flow analysis is a subset of dataflow analysis that helps to statically reason about the dependencies among program constructs such as variables and memory blocks.
It underpins many crucial program analysis techniques that are widely used in compiler optimizations and for finding security vulnerabilities: points-to analysis~\cite{zhao2018parallel}, null pointer analysis~\cite{xu2019vfix,ma2015practical}, and taint analysis~\cite{grech2017p}, to name a few.
In order to operate effectively, such techniques usually require precise value-flow information that is context- and flow-sensitive as well as interprocedural~\cite{wangGraspanSinglemachineDiskbased2017}.

Seminal work by Reps et al.~\cite{reps1995ifds} demonstrated that interprocedural context- and flow-sensitive value-flow analysis can be expressed as reachability between nodes within a program graph.
Their algorithmic framework \textit{IFDS} operates on a graph structure in which each node maps to a value-flow statement within the target program.
But, while solved in theory,
in practice the adoption of static value-flow frameworks is still limited by severe scalability issues.
In fact, as we will show in Section~\ref{sec:eval}, currently there exists no static analysis framework that is able to determine precise value-flows for larger real world codebases such as OpenSSL and FFmpeg on single commodity PCs.

The memory and runtime requirements of static value-flow analysis frameworks such as IFDS are largely governed by the graph representation of the program
and the performance of the graph reachability algorithm.
Indeed, previous approaches to solving the scalability problem have often focused on working around the shortcomings of established graph-reachability
algorithms~\cite{hePerformanceBoostingSparsificationIFDS2019,liScalingIFDSAlgorithm2021,arztSustainableSolvingReducing2021}.
In contrast, in this paper, we present a solution that is based on a novel sparse graph representation, leading to a reachability algorithm that is highly efficient in answering value-flow queries.


Our key insight is that program graphs representing value-flow can be constructed to have a low tree width~\cite{ROBERTSON198449}, which is a measure of how similar a graph is to a tree.
We found that performing a depth-first traversal in the \emph{opposite} direction of SSA def-use chains will result in a graph with a significantly reduced number of \textit{non-tree} edges.
Based on this insight, we develop a novel graph reachability algorithm based on tree traversal which significantly reduces processing time and memory requirements compared to previous approaches.
In fact, our new algorithm allows us to determine dependencies between two \textit{arbitrary} instructions in constant time for most queries.
Additionally, the resulting graph representation is \textit{sparse}, thereby further reducing processing time and memory requirements by
incorporating only program statements relevant for value-flow propagation.

We implemented these ideas in a framework that we call \coolname{}\footnote{The name \coolname{} pays homage to IFDS, while our reversal of the letters alludes to the fact that our technique processes nodes in the reverse order.}.
To quantify the improved performance of our approach, we evaluate \coolname{} against two state-of-the art static value-flow analysis frameworks: Phasar~\cite{schubertPhASARInterproceduralStatic2019} and SVF~\cite{sui2016svf}.
Phasar is a popular framework implementing the IFDS algorithm,
while SVF presents a different approach to compute precise static value-flow information based on def-use chains that are constructed based on pre-computed points-to information.
As our evaluation shows, \coolname{} is able to scale to significantly larger codebases on commodity hardware than previous solutions,
using significantly less memory overall, and running significantly faster.
Therefore, \coolname{} constitutes the first static solution to resolve interprocedural context- and flow-sensitive value-flow
for real-world software projects, reaching to over a million lines of code, under realistic resource limitations.
In summary our contributions are the following:
\begin{itemize}
\item an extension of LLVM IR containing two new operations that significantly simplify the modeling of value-flows in the presence of pointer aliasing,
\item a resulting lightweight and precise sparse representation of def-use chains for value-flow analysis incorporating top-level and address-taken variables,
\item an algorithm to solve value-flows via graph reachability that scales almost linearly with the number of processed vertices, requires significantly less memory and is significantly faster than previous solutions, and
\item a full source-language agnostic implementation of our technique based on LLVM IR that scales to large real-world projects containing hundreds of thousands of lines of code.
\end{itemize}
We further pledge to make the source code of our project freely available under an open-source license.

\section{Background}
In this Section, we explain terminologies and concepts that will serve as the building blocks for rest of the paper.

\subsection{Flow-Sensitive Value-Flow Analysis}

Value-flow analysis models the specific propagation of data through a program's storage locations.
A flow-sensitive analysis respects the program's control flow and calculates results for each program point. In contrast, flow-insensitive analysis ignores statement ordering and computes a single solution that is sound for all program points. Traditionally, to achieve flow-sensitivity, value-flow analysis is built on top of a monotone dataflow analysis framework~\cite{kam1977monotone}. The analysis follows control flows in the control-flow graph (CFG), while accounting for changes in dataflow facts made by each statement.
For example, in a null pointer analysis, a dataflow fact is a set of potentially-null variables at a given program point.
For each statement $i$, the transfer function calculates two sets: $IN_i$ and $OUT_i$, which represent the sets of dataflow facts that hold right before and right after statement $i$. In the context of a null pointer analysis, the transfer function might add (or subtract) variables that are (or not) potentially-null. $IN_i$ is equal to $OUT_{i-1}$ passed down from previous statement in the control flow, in the case of having multiple predecessor statements, $IN_i$ is the set union of $OUT_p, \forall p \in predecessors(i)$. The algorithm repeatedly performs such updates on every statement until it reaches a global fixed point.

This approach usually comes with a high performance overhead. Dataflow facts are propagated to every program point, despite the fact that only a small portion of them is needed to answer the dataflow query we are interested in. In addition, maintaining two sets per statement tends to create a large memory footprint as well.

\subsubsection*{SSA-Based Sparse Value-Flow Analysis}\label{sec:background-ssa-sparse}
In recent years, sparse value-flow analysis~\cite{hardekopf2009semi,sui2016svf,oh2012design,hardekopf2011flow} has introduced a promising solution to the aforementioned problem.
Sparse value-flow analysis avoids propagating dataflow facts through program statements that are unrelated to the analysis. A common way is to perform the analysis on Static-Single Assignment (SSA) form~\cite{cytron1991efficiently} of programs.

In SSA, each variable is defined exactly once in a ``static'' view of the program, i.e., disregarding ``dynamic'' reassignments of variables that may happen at the same program location inside of loops. If there are multiple static definitions of a variable in the original program (such as multiple separate program locations that perform assignments to the variable), then each such assignment of the original program variable becomes a separate SSA variable or value. SSA form expresses value definitions and usage information, also called def-use chains, explicitly, so that tracing values to their immediate uses becomes much easier. Therefore, def-use chains can help us to rule out irrelevant data flows and greatly improve the efficiency of value-flow analyses~\cite{reif1977symbolic}.

Tracking every variable in SSA form is difficult as it has to account for potential aliasing among pointer variables. To tackle this problem, mainstream compilers such as GCC~\cite{novillo2004design} and LLVM~\cite{lattner2004llvm} adopt a variant of SSA called \textit{partial} SSA which divides variables into two categories: top-level and address-taken. Top-level variables cannot be referenced indirectly via a pointer and can be trivially converted into SSA form; address-taken variables are referenced indirectly via top-level pointer variables and are not represented in SSA form at all. This additional layer of indirection makes it more difficult for us to track value-flows on address-taken variables in partial SSA. In Section~\ref{sec:preprocessing}, we will show our solution to this problem.

\subsection{Value-Flow Analysis as Graph Reachability}\label{sec:background-graph-reachability}
Reps et al. pioneered the idea of solving traditional dataflow analyses by turning them into graph reachability problems in their IFDS framework~\cite{reps1995ifds}, inspiring a whole body of research on graph reachability based program analysis~\cite{vardoulakis2010cfa2,kodumal2004set,bastani2015specification,melski2000interconvertibility}.
In IFDS, dataflow analysis is performed on an \textit{exploded supergraph}, in which each vertex represents a dataflow fact at a specific program point. Each statement has a local transfer function that describes the vertex (i.e. dataflow facts) mapping before and after the statement. Multiple such edges transitively compose to a single \textit{PathEdge}. Thus, an intraprocedural dataflow query can boil down to answering whether the vertex of a dataflow source reaches the program points we are interested in via a PathEdge, which implies flow-sensitivity. Interprocedurally, the PathEdge are extended from a call site to a callee and back to the same call site. Context-sensitivity is achieved 
by matching call and return edges.

\subsection{ 
Reachability via Depth-First Tree Intervals}\label{sec:background-dft-intervals}
\begin{figure}
    \centering
        \includegraphics[clip, trim=3cm 4.2cm 5.5cm 4.2cm, width=.7\columnwidth]{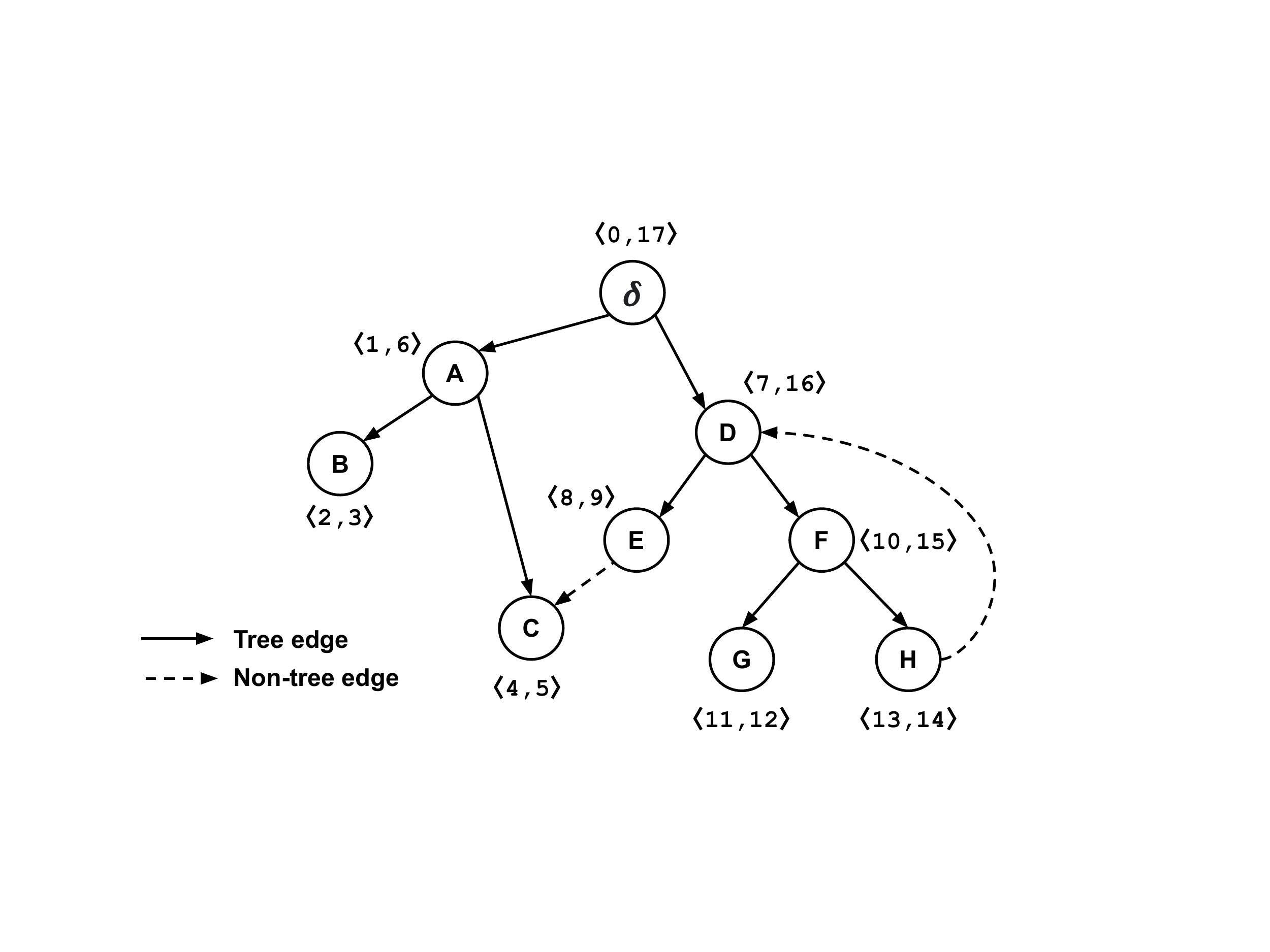}
    \caption{A graph annotated with DFT intervals}
    \label{fig:dft-intervals}
\end{figure}

A \emph{Depth-First Tree (DFT)} is a common approach to compute graph reachability. A DFT is an ordered spanning tree derived from the process of \emph{Depth-First Search (DFS)}~\cite{cormen2022introduction}.
Each vertex of a DFT is assigned an integer interval $\interval{s_v, e_v}$:
$s_v$ is the discovery timestamp when $v$ is first visited and
$e_v$ is the finish timestamp when all out-neighbors of $v$ have been visited.
The timestamp is initialized with zero and it is increased by one upon visiting a new out-neighbor from a vertex.
Formally speaking, an interval has the following invariant:
$$
s_v, e_v \in \mathbb{Z}, s_v \geq 0 \land e_v \geq 0 \land e_v > s_v
$$
Figure~\ref{fig:dft-intervals} shows an example graph with its DFT intervals. To simplify the problem without losing generality, we always add a pseudo vertex $\delta$ to connect every vertex in the graph, such that the spanning tree has a single root.

For a given vertex in the spanning tree, its interval always \emph{subsumes} the intervals of all vertices in its subtree. An interval $\interval{s_k, e_k}$ is said to subsume another interval $\interval{s_l, e_l}$, denoted by $\interval{s_k, e_k} \supseteq \interval{s_l, e_l}$, if they have the following properties:
$$
s_k \leq s_l \land e_k \geq e_l \implies \interval{s_k, e_k} \supseteq \interval{s_l, e_l}
$$

Vertex $F$ in Figure~\ref{fig:dft-intervals}, for example, has an interval of $\interval{10,15}$, which subsumes the intervals of its children $\interval{11,12}$ and $\interval{13,14}$. Conversely, vertex $E$ and $F$ are siblings so none of their intervals subsumes one or the other. We distinguish between tree edges (solid lines) and non-tree edges (dashed lines) in Figure~\ref{fig:dft-intervals}. There are two types of non-tree edges: cross edge and back edge. They are defined using the interval relationship: for vertices $k$ and $l$, alone with their corresponding intervals $\interval{s_k, e_k}$ and $\interval{s_l, e_l}$, Edge $k \to l$ is
\begin{itemize}
    \item a cross edge if $e_l < s_k \lor e_k < s_l$, and
    \item a back edge if $\interval{s_l, e_l} \supseteq \interval{s_k, e_k}$.
\end{itemize}
For example, in Figure~\ref{fig:dft-intervals}, edge $E \to C$ is a cross edge, and edge $H \to D$ is a back edge.

By comparing the intervals of arbitrary two vertices, using the subsuming relation defined earlier, we can easily know whether they can reach each other through (spanning) tree edges in \textit{constant time}. However, this property only holds for tree edges.
For reachability queries on generic graphs, we also need to consider paths that go through non-tree edges. To answer those queries, several previous works~\cite{wang2006dual,yildirim2010grail,he2005compact} have proposed a variety of ad-hoc solutions on top of DFT intervals. In Section~\ref{sec:design-augmented-reachability}, we are going to introduce an augmented DFT-interval graph reachability algorithm to solve this problem.


\section{\coolname{} Design and Implementation}
\coolname{} is designed to be programming-language agnostic on its input program and can process any program compiled into LLVM IR. Section~\ref{sec:preprocessing} will cover necessary preprocessing to convert 
this initial LLVM IR
into a form more favorable to value-flow analysis. Sections~\ref{sec:design-structure-n-workflow} to~\ref{sec:design-interprocedural} will cover the details of our main algorithm and implementation.

\subsection{Preprocessing}\label{sec:preprocessing}
\coolname{} relies heavily on the sparse SSA program representation to track value-flows efficiently.
However, as discussed in Section~\ref{sec:background-ssa-sparse}, the partial SSA form used by LLVM IR excludes address-taken variables.
To track the value-flows of address-taken variables we convert the original input program into a custom representation that improves handling of indirect memory operations and pointers.

We implement this custom program representation using MLIR~\cite{lattner2021mlir}, a versatile framework to create custom intermediate representations (IRs) for program analyses and compiler optimizations.
A custom IR is called \emph{dialect} in MLIR. It consists of a type system and various building blocks that define the semantics like operation, block, region, and attribute, to name a few. A program statement or LLVM IR instruction is usually modeled by a MLIR operation.
We create our own \coolname{} dialect on top of existing LLVM IR constructions with two custom operations: \texttt{dfi.store} and \texttt{dfi.call}.

\subsubsection*{\texttt{dfi.store} operation}\label{sec:background-defi-store}
A \texttt{dfi.store} operation has exactly the same run-time behavior as an LLVM \texttt{store} instruction. However, unlike a normal \texttt{store} instruction that doesn't produce a result, \texttt{dfi.store} produces a pointer as the result. The result pointer has the same aliasing characteristics as the destination pointer operand, \emph{i.e.}, the memory address to which the value is stored.

\texttt{dfi.store} essentially applies the same SSA renaming used by LLVM on pointer variables and adds \textit{explicit} def-use chains for strong updates to an address-taken variable. More specifically, upon each strong memory store, we rename the original pointer to the result pointer produced by \texttt{dfi.store}.

Listing~\ref{lst:defi-store-before} shows a simple MLIR snippet consisting of memory load and store operations on function argument \texttt{\%p}. Listing~\ref{lst:defi-store-after}, on the other hand, is the same snippet with \texttt{llvm.store} replaced by \texttt{dfi.store}.
Take \texttt{llvm.load} operation on line~3 of Listing~\ref{lst:defi-store-before} as an example: before the replacement, it was loading content from \texttt{\%p}. But on the same line of Listing~\ref{lst:defi-store-after}, we can see that it's now loading content from the result pointer \texttt{\%p0} of \texttt{dfi.store} from the previous line.

The presence of \texttt{dfi.store} operations dramatically simplifies the value-flow propagations on address-taken variables. We can now track memory dependencies w.r.t.~strong updates in the same way as tracking scalar value-flows. It is similar to the idea proposed by Chow et al.~\cite{chow1996effective} and MemorySSA~\cite{novillo2007memoryssa}. In contrast to previous works, \texttt{dfi.store} adopts a lightweight design and excludes weak updates on address-taken variables (\emph{i.e.} \textit{MayDef} and \textit{MayUse}), which are more expensive to compute, from its representation. Nevertheless, a client analysis is capable of accounting for those weak memory dependencies with the flexible framework provided by \coolname{}.
\begin{lstlisting}[
    style=MLIRStyle,
    captionpos=b, caption={IR before conversion to \texttt{dfi.store}},
    label={lst:defi-store-before}]
llvm.func @f(%v: i32, %p: !llvm.ptr<i32>) {
    llvm.store %v, ?\colorbox{green}{\%p}? : !llvm.ptr<i32>
    %v0 = llvm.load ?\colorbox{green}{\%p}? : !llvm.ptr<i32>
    llvm.store %v0, ?\colorbox{green}{\%p}? : !llvm.ptr<i32>
    %v1 = llvm.load ?\colorbox{green}{\%p}? : !llvm.ptr<i32>
}
\end{lstlisting}
\vspace*{-0.5em}
\begin{lstlisting}[
    style=MLIRStyle,
    captionpos=b, caption={IR after conversion to \texttt{dfi.store}},
    label={lst:defi-store-after}]
llvm.func @f(%v: i32, %p: !llvm.ptr<i32>) {
    ?\colorbox{yellow}{\%p0}? = dfi.store %v, ?\colorbox{green}{\%p}? : !llvm.ptr<i32>
    %v0 = llvm.load ?\colorbox{yellow}{\%p0}? : !llvm.ptr<i32>
    ?\colorbox{pink}{\%p1}? = dfi.store %v0, ?\colorbox{yellow}{\%p0}? : !llvm.ptr<i32>
    %v1 = llvm.load ?\colorbox{pink}{\%p1}? : !llvm.ptr<i32>
}
\end{lstlisting}
\vspace*{-0.5em}
\subsubsection*{\texttt{dfi.call} operation}\label{sec:background-defi-call}
To support interprocedural analysis, it is essential to capture value-flows of the callee function at a given call site. For callees that have no side effects (e.g. pure functions), it suffices to propagate value-flows through function arguments and the return value. However, we also need to consider \textit{output arguments} where results are carried through pointer type function parameters. In order to capture the value-flows propagated through output arguments, \coolname{} replaces normal function call operations, \texttt{llvm.call}, with custom \texttt{dfi.call} operations.

Similar to \texttt{dfi.store}, for each pointer argument in the original \texttt{llvm.call}, we add a pointer-type result in the result list of \texttt{dfi.call}. Listing~\ref{lst:defi-call-after} and~\ref{lst:defi-call-before} in the Appendix~\ref{sec:appendix:listings} show the result of replacing \texttt{llvm.call} with \texttt{dfi.call}.
\footnote{For better readability, we will omit the type notation of all \texttt{dfi.call} occurrences in rest of the paper.}



\subsection{Analysis Engine}\label{sec:design-structure-n-workflow}
\begin{figure}
    \centering
        \includegraphics[clip, trim=0.5cm 4cm 0.6cm 4cm, width=.9\columnwidth]{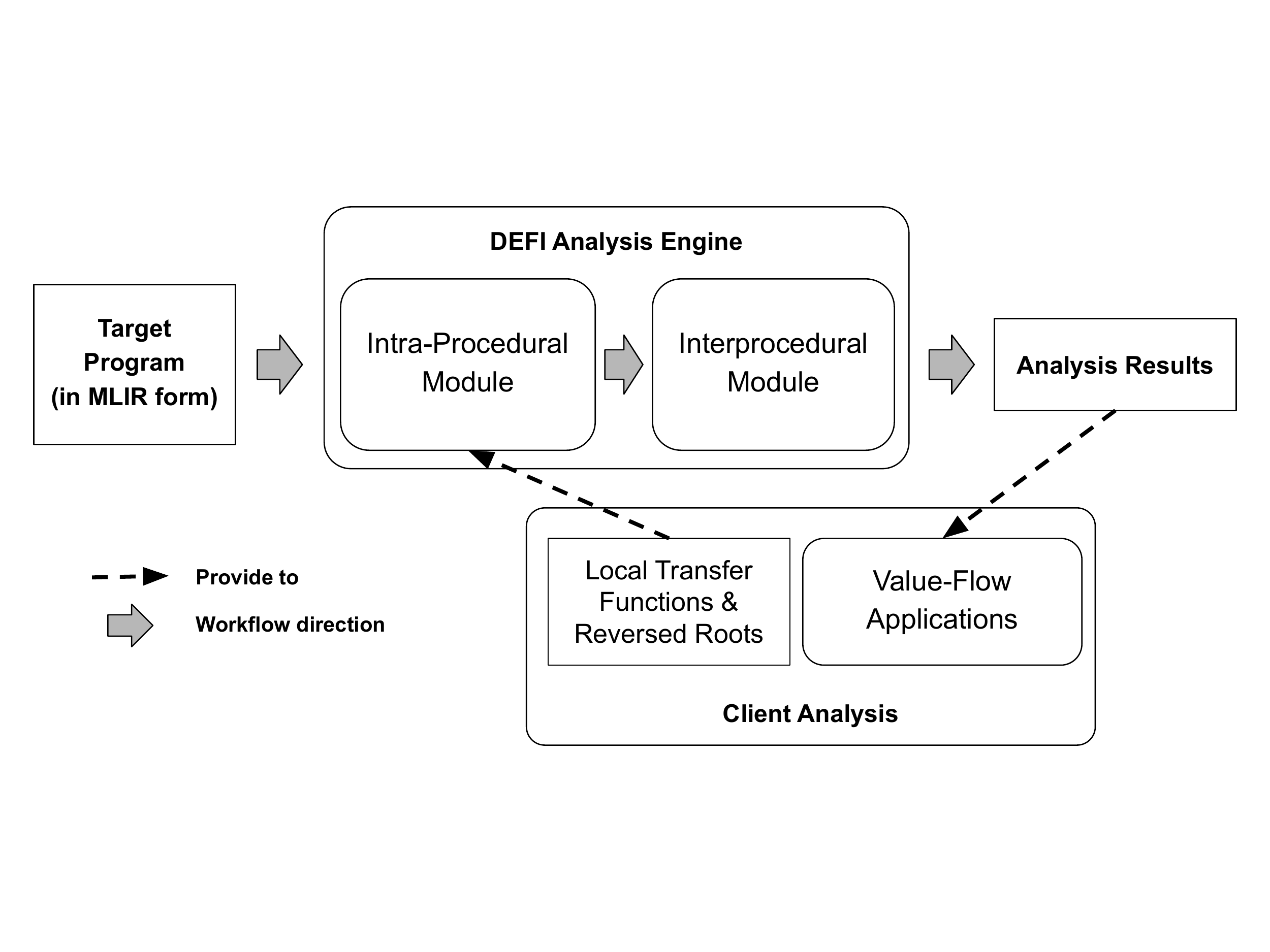}
    \caption{Structure of the analysis engine in \coolname{}}
    \label{fig:defi-structure}
\end{figure}

Figure~\ref{fig:defi-structure} shows the overall structure and workflow in \coolname{} after the preprocessing. \coolname{} consists of two primary components: the main analysis engine and the client analysis.

First, as further detailed in Section~\ref{sec:design-intraprocedural}, intraprocedural DFT intervals are computed by the analysis engine based on value-flow mapping information provided by the client analysis. Next, as described in Section~\ref{sec:design-interprocedural}, the intraprocedural value-flows will be propagated to their callers in order to support interprocedural queries. Finally, the client analysis will receive \textbf{flow-} and \textbf{context-sensitive} value-flow results to solve its domain-specific applications.

\subsection{Intra-procedural Analysis}\label{sec:design-intraprocedural}
This Section discusses the algorithm of tracking intraprocedural value-flows. As mentioned in Section~\ref{sec:background-graph-reachability}, value-flow analysis can be boiled down to a graph reachability problem, where each vertex is a dataflow fact on a certain program point and value-flow queries are answered by determining the reachability of two vertices.

Similar to IFDS, \coolname{} also approaches the intraprocedural part of this problem by employing a local transfer function for each statement and a graph reachability solving technique. However, instead of the tabulation-based strategy used by the original IFDS technique, \coolname{} adopts a novel graph reachability solving technique based on depth-first spanning tree intervals introduced in Section~\ref{sec:background-dft-intervals}.
The idea is that for a given value-flow analysis problem, if every relevant operation in the program is annotated with DFT intervals, we are able to determine the reachability between two arbitrary operations by comparing their intervals in nearly constant time.

\begin{figure}
    \centering
        \includegraphics[clip, trim=3cm 5.2cm 3.3cm 5.2cm, width=.8\columnwidth]{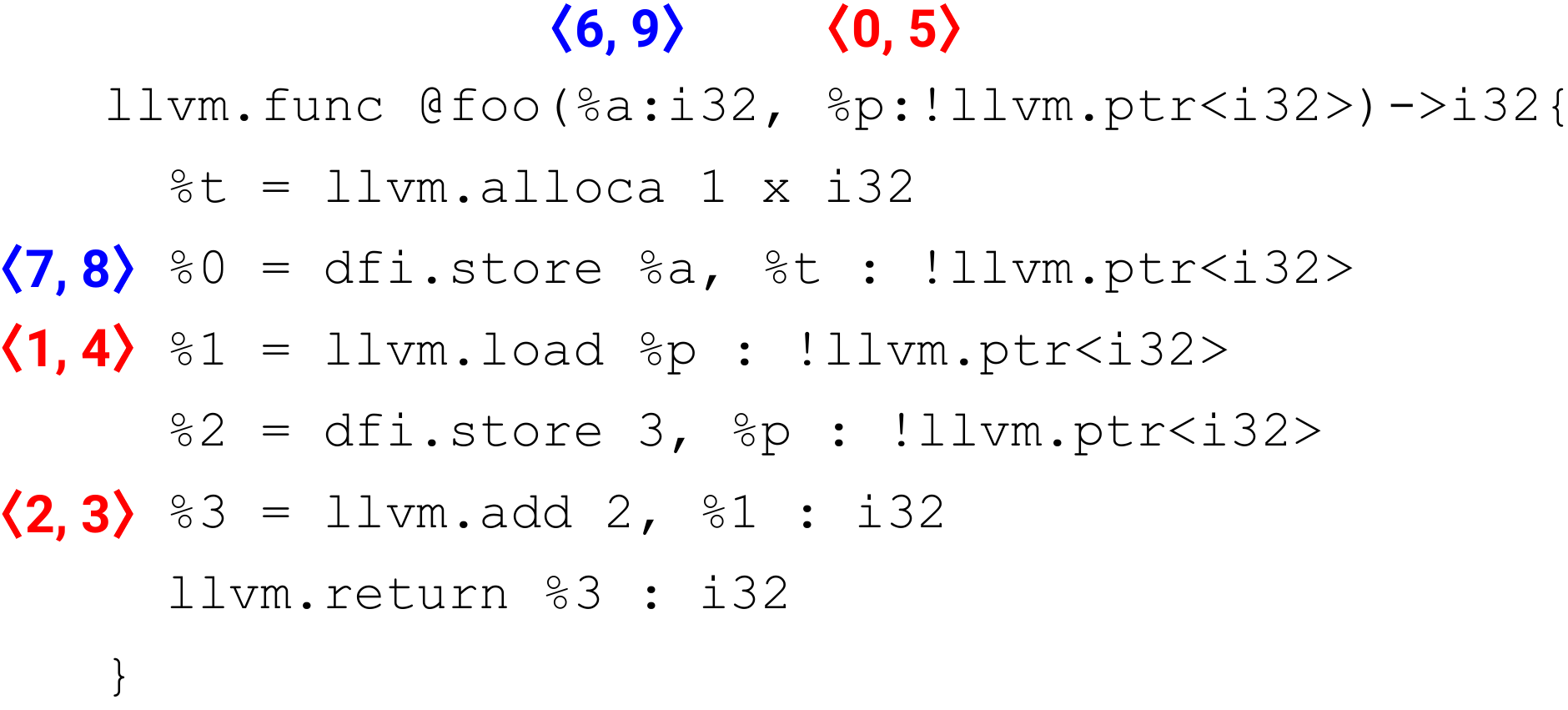}
    \caption[Caption for LOF]{DFT intervals in a MLIR function\footnotemark}
    \label{fig:defi-interval-demo}
\end{figure}
\footnotetext{To simplify the code, we "inline" the constant operands in both \texttt{llvm.add} operations. Normally, those constants will be defined by a dedicated operation, \texttt{llvm.constant}.}

To demonstrate this concept, Figure~\ref{fig:defi-interval-demo} shows a MLIR function in which the values are annotated with DFT intervals. For instance, value $\%p$ and $\%1$ have $\interval{0,5}$ and $\interval{1,4}$ for their intervals, respectively. These intervals are annotated on the spanning trees derived from the \textbf{SSA def-use graph}, consisting of edges that go from a single SSA value definition to its uses.

Assume a \textit{taint analysis} is performed on Figure~\ref{fig:defi-interval-demo} and $\%p$ is a tainted value. To determine if the return value $\%3$ is tainted it is sufficient to check if its interval $\interval{2,3}$ is subsumed by that of $\%p$ $\interval{0,5}$. In this case, since $\interval{0,5} \supseteq \interval{2,3}$ per our definition in Section~\ref{sec:background-dft-intervals}, the return value is tainted. In fact, both $\%1$ and $\%3$ are tainted, as their intervals are both inside the subtree of $\interval{0,5}$. On the other hand, $\%0$, whose interval is $\interval{7,8}$, is not tainted because it resides in a different DFT subtree rooted at $\interval{6,9}$.

To calculate the trees and associated intervals shown in Figure~\ref{fig:defi-interval-demo} the client analysis customizes the traversal mechanism introduced in Section~\ref{sec:background-dft-intervals}. First, for each function, the client analysis chooses a set of root vertices to start the traversal. Note that the traversal timestamp in each function always starts from zero.
Second, for each vertex in the function, a local transfer function is provided by client analysis to dictate the out-going vertices to visit next. In the context of taint analysis on Figure~\ref{fig:defi-interval-demo}, the client analysis picks $\%a$ and $\%p$ as the traversal roots. For the local transfer function, the rules for each kind of operation are shown in Figure~\ref{fig:defi-local-transfer-function}. Take the rule for \texttt{dfi.store} as an example, in which the result pointer $\%q$ is tainted only if the stored value $\%v$ is tainted. It effectively stops the DFT traversal to go from $\%p$ to $\%q$, but allows the path from $\%v$ to $\%q$.
\begin{figure}
    \centering
    \resizebox{\columnwidth}{!}{%
    \begin{tabular}{|c|c|c|}
       \hline
\texttt{\%q = dfi.store \%v, \%p}
           &
       \texttt{\%r = llvm.add \%a, \%b}
           &
       \texttt{\%v = llvm.load \%p}
           \\
       \hline
           \raisebox{-0.5\height}{\usetikzlibrary{arrows}
\tikzset{>=latex}
\begin{tikzpicture}
    \node[font=\small] (A) at (0,1) {\texttt{\%v}};
    \node[font=\small] (B) at (1,1) {\texttt{\%p}};
    \node[font=\small] (C) at (0,0) {\texttt{\%q}};

    \path [->,thick] (A) edge node[right,font=\small] {} (C);
\end{tikzpicture}} &
           \raisebox{-0.5\height}{\usetikzlibrary{arrows}
\tikzset{>=latex}
\begin{tikzpicture}
    \node[font=\small] (A) at (0,1) {\texttt{\%a}};
    \node[font=\small] (B) at (1,1) {\texttt{\%b}};
    \node[font=\small] (C) at (0,0) {\texttt{\%r}};

    \path [->,thick] (A) edge node[right,font=\small] {} (C);
    \path [->,thick] (B) edge node[right,font=\small] {} (C);
\end{tikzpicture}} &
           \raisebox{-0.5\height}{\usetikzlibrary{arrows}
\tikzset{>=latex}
\begin{tikzpicture}
    \node[font=\small] (A) at (0,1) {\texttt{\%p}};
    \node[font=\small] (B) at (0,0) {\texttt{\%v}};

    \path [->,thick] (A) edge node[right,font=\small] {} (B);
\end{tikzpicture}} \\
           \hline
    \end{tabular}}
    \caption[Caption for LOF]{Local taint analysis transfer functions on Figure~\ref{fig:defi-interval-demo}}
    \label{fig:defi-local-transfer-function}
\end{figure}

The DFT interval scheme described in the previous example only details the computation of reachability based on (spanning) \textit{tree} edges. In the following Section we will discuss a more general reachability problem w.r.t both \textit{tree and non-tree} edges.

\subsubsection*{Non-Tree Edges in SSA Def-Use Graph}\label{sec:design-non-tree-edges}
Non-tree edges inhibit the application of the simple interval-based graph reachability algorithm on generic SSA def-use graphs. To generalize our solution over those graphs, we first discuss program constructs that result in non-tree edges.
\begin{lstlisting}[
    style=MLIRStyle,
    captionpos=b, caption={A MLIR function},
    label={lst:defi-interval-example2}]
llvm.func @f(%a: i32, %b: i32, %c: i32) -> i32 {
    %t = llvm.add %a, %c : i32
    %r = llvm.mul 3, %t : i32
    llvm.return %r : i32
}
\end{lstlisting}
\vspace*{-0.5em}
Figure~\ref{fig:defi-interval-demo2} shows the SSA def-use graph for Listing~\ref{lst:defi-interval-example2}. In Figure~\ref{fig:defi-interval-demo2}, each vertex is labeled with an operation or value and edges are labeled with the value being used. Starting the creation of DFT intervals from function argument $\%a$ will produce three edges: $\%a$, $\%t$, and $\%r$. However, when we proceed to visit rest of the vertices, the edge between $\%c$ and $llvm.add$ becomes a non-tree edge, since the latter operation has been visited before. We observe that, with this scheme, a non-tree edge appears if there is more than one operand in an operation. The additional operands create multiple parent vertices for the operation resulting in non-tree edges. Therefore, the number of non-tree edges is proportional to the number of operands. Unfortunately, the majority of nodes in standard LLVM IR have more than one operand.

\begin{figure}[]
    \centering
    \begin{subfigure}{.5\columnwidth}
        \centering
        \tikzset{>=latex}
\begin{tikzpicture}
    \node[rectangle,draw,font=\tiny] (A) at (0,2.25) {\texttt{\%a}};
    \node[rectangle,draw,font=\tiny,fill=lightgray2] (B) at (0.8,2.25) {\texttt{\%c}};
    \node[rectangle,draw,font=\tiny] (C) at (0.0,1.5) {\texttt{\%t = llvm.add \%a, \%c}};
    \node[rectangle,draw,font=\tiny] (D) at (0.0,0.75) {\texttt{\%r = llvm.mul 3, \%t}};
    \node[rectangle,draw,font=\tiny] (E) at (0.0,0) {\texttt{llvm.return \%r}};

    \path [->] (A) edge node[left,font=\tiny]  {\texttt{\%a}} (C);
    \path [->] (B) edge node[right,font=\tiny] {\texttt{\%c}} (C);
    \path [->] (C) edge node[right,font=\tiny] {\texttt{\%t}} (D);
    \path [->] (D) edge node[right,font=\tiny] {\texttt{\%r}} (E);
\end{tikzpicture}
        \caption{SSA def-use graph}
        \label{fig:defi-interval-demo2}
    \end{subfigure}%
    \begin{subfigure}{.5\columnwidth}
        \centering
        \tikzset{>=latex}
\begin{tikzpicture}
    \node[rectangle,draw,font=\tiny] (A) at (0,0) {\texttt{\%a}};
    \node[rectangle,draw,font=\tiny] (B) at (1,0) {\texttt{\%c}};
    \node[rectangle,draw,font=\tiny] (C) at (0.5,0.75) {\texttt{\%t = llvm.add \%a, \%c}};
    \node[rectangle,draw,font=\tiny] (D) at (0.5,1.5) {\texttt{\%r = llvm.mul 3, \%t}};
    \node[rectangle,draw,font=\tiny] (E) at (0.5,2.25) {\texttt{llvm.return \%r}};

    \path [->] (E) edge node[right,font=\tiny] {\texttt{\%r}} (D);
    \path [->] (D) edge node[right,font=\tiny] {\texttt{\%t}} (C);
    \path [->] (C) edge node[right,font=\tiny] {\texttt{\%c}} (B);
    \path [->] (C) edge node[left,font=\tiny]  {\texttt{\%a}} (A);
\end{tikzpicture}
        \caption{Reversed SSA def-use graph}
        \label{fig:defi-reversed-dft}
    \end{subfigure}
    \caption{SSA def-use graphs for Listing~\ref{lst:defi-interval-example2}}
\end{figure}

\subsubsection*{Reversed DFT Traversal}\label{sec:design-reserved-dft-traversal}
To effectively reduce the number of non-tree edges in a SSA def-use graph, \coolname{} adopts a novel solution: traversing the graph in the \emph{reverse direction} when building DFT intervals. In Figure~\ref{fig:defi-interval-demo2} the traversal direction goes from a SSA definition to its uses. If we reverse this direction and go from a SSA use to its value definition, as demonstrated in Figure~\ref{fig:defi-reversed-dft}, the number of non-tree edges no longer depends on the fixed number of operands. Instead, for a given value, the number of non-tree edges is proportional to the number of its SSA \textit{uses}.

This led us to an important new realization: in most of the real-world codebases, the majority of the SSA values have a \textit{single} or even \textit{no SSA use}. Table~\ref{tab:ssa-num-value-users} in the Appendix shows the percentage of SSA values against different number of SSA uses in 6 real-world codebases. The statistics show that \textbf{80\%} to \textbf{90\%} of the SSA values have a single or no use in all benchmarks. In other words, if we traverse the SSA def-use graphs in the opposite direction, the resulting number of non-tree edges is far smaller than the number of tree edges.
\vspace*{-0.5em}
\paragraph{Reversed DFT Root}
Reversed DFT traversal starts from a set of vertices called reversed roots. They are the out-neighbors of pseudo vertex $\delta$ introduced in Section~\ref{sec:background-dft-intervals}. In \coolname{}, the client analysis can pick its own reversed roots to begin the traversal. Since they have an implication on the set of vertices to be visited, it is worth discussing the intuition behind reversed roots in the context of value-flow analysis.

In a backward analysis, a reversed root naturally assumes the role of value-flow \textit{origin}. For example, for a query asking if value $x$ is live on operation $I$, a classic live-variable analysis, we can pick values used by $I$ (\emph{i.e.} its operands) as reversed roots.
On the other hand, reversed roots can be seen as all possible \textit{destinations} in a forward analysis setup. For instance, in the taint analysis that asks if sensitive information (\emph{i.e.} tainted values) flows to any system call, we can pick all system call sites as the reversed roots such that all possible value-flow paths can be considered. It is worth noting that \coolname{} is able to consider all possible paths at once because our algorithm is both time and space efficient, which we will show in Section~\ref{sec:eval}.

\subsubsection*{Augmented DFT-Interval Graph Reachability}\label{sec:design-augmented-reachability}
With reduced number of non-tree edges, \coolname{} augments the DFT-interval-based algorithm to solve graph reachability in reversed SSA def-use graphs. In short, this method \textit{duplicates} the interval upon encountering a non-tree edge, such that we can use a similar subsuming relationship between two intervals to determine their reachability.

To support our method, we introduce a new data structure: interval set. An interval set $\Pi$ is a collection of intervals
$$
\{\interval{s_1, e_1}, \interval{s_2, e_2}, ..., \interval{s_n, e_n}\}
$$
in which every element separates themselves with each other by least one timestamp. \emph{i.e.}
$$
\forall \interval{s_i, e_i}, \interval{s_j, e_j} \in \Pi, i \neq j \implies e_i < s_j - 1 \vee e_j < s_i - 1
$$
An interval set $\intervalset{i}$ is said to subsume another set $\intervalset{j}$, namely $\intervalset{i} \supseteq \intervalset{j}$, if any of the interval in $\intervalset{i}$ can subsume another interval in $\intervalset{j}$. \emph{i.e.}
$$
\exists \interval{s_k, e_k} \in \intervalset{i}, \interval{s_l, e_l} \in \intervalset{j} \emph{ s.t. } \interval{s_k, e_k} \supseteq \interval{s_l, e_l}
$$
Two interval sets $\intervalset{i}$ and $\intervalset{j}$ can be merged by operator $\cup$, denoted as $\intervalset{i} \cup \intervalset{j}$.
Let $\intervalset{k}$ be the merged interval set from $\intervalset{i}$ and $\intervalset{j}$, it can subsume both $\intervalset{i}$ and $\intervalset{j}$ \emph{i.e.} $\intervalset{k} \supseteq \intervalset{i} \wedge \intervalset{k} \supseteq \intervalset{j}$.
In our augmented DFT-interval-based reachability algorithm, each SSA def-use graph vertex $v$ is associated with an interval set $\intervalset{v}$ (rather than a single interval). Vertex $v_i$ can reach $v_j$ if and only if $\intervalset{v_j}$ subsumes $\intervalset{v_i}$. \emph{i.e.}
$$
v_i \leadsto v_j \iff \intervalset{v_j} \supseteq \intervalset{v_i} \land \intervalset{v_i} \neq \emptyset \land \intervalset{v_j} \neq \emptyset
$$
Note that since we build DFT intervals in \textit{reversed} direction, a vertex can reach another vertex if the interval set of the destination subsumes that of the source. 
To build interval sets for each vertex in the graph, the interval set of each vertex is first initialized with a single interval created from a normal DFT-interval building process (see Section~\ref{sec:background-dft-intervals}).
Next, non-tree edges are incorporated.

\begin{figure}
    \centering
    \begin{subfigure}{.4\columnwidth}
        \centering
            \includegraphics[clip, trim=6.5cm 5.2cm 6.2cm 5.5cm, width=\linewidth]{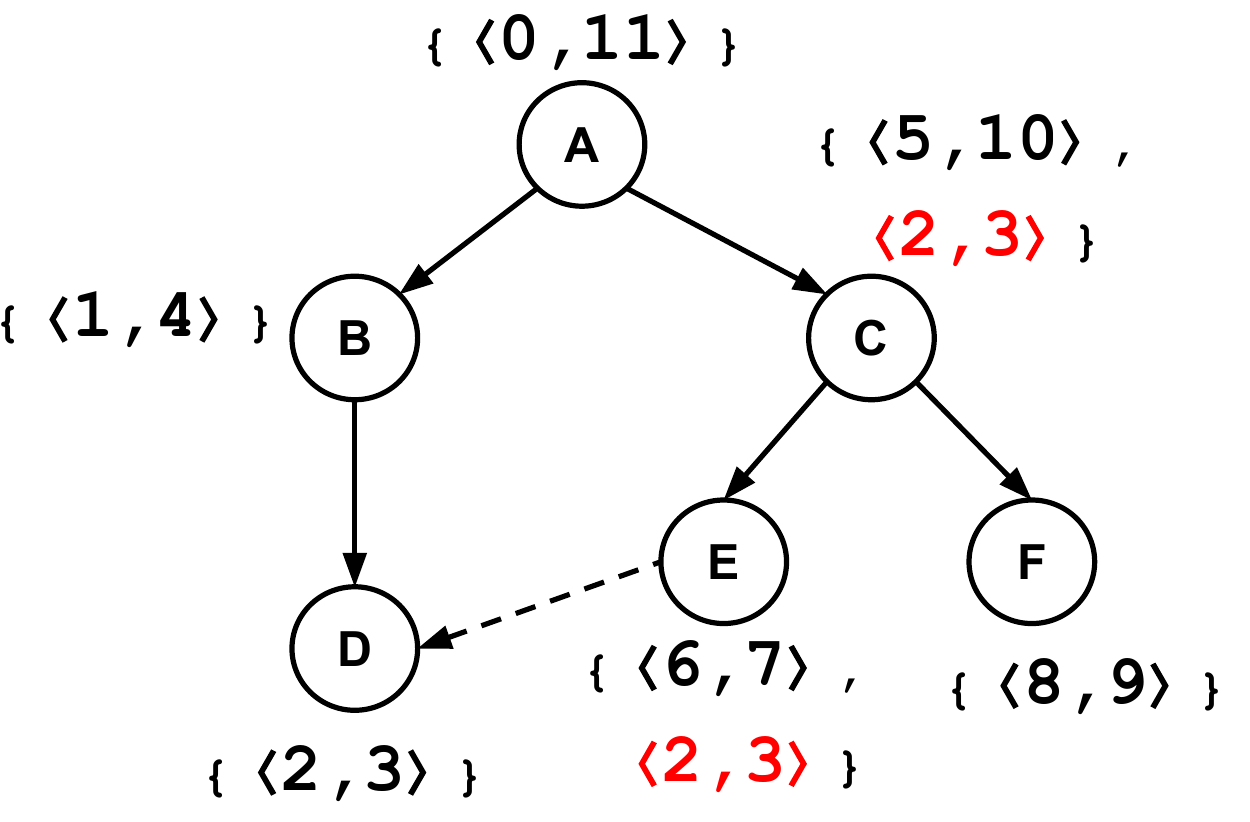}
        \caption[Caption for LOF]{Cross Edge}
        \label{fig:dft-interval-cross-edge}
    \end{subfigure}%
    \begin{subfigure}{.6\columnwidth}
        \centering
        \frame{
            \includegraphics[clip, trim=2.2cm 5cm 3.0cm 5cm, width=\linewidth]{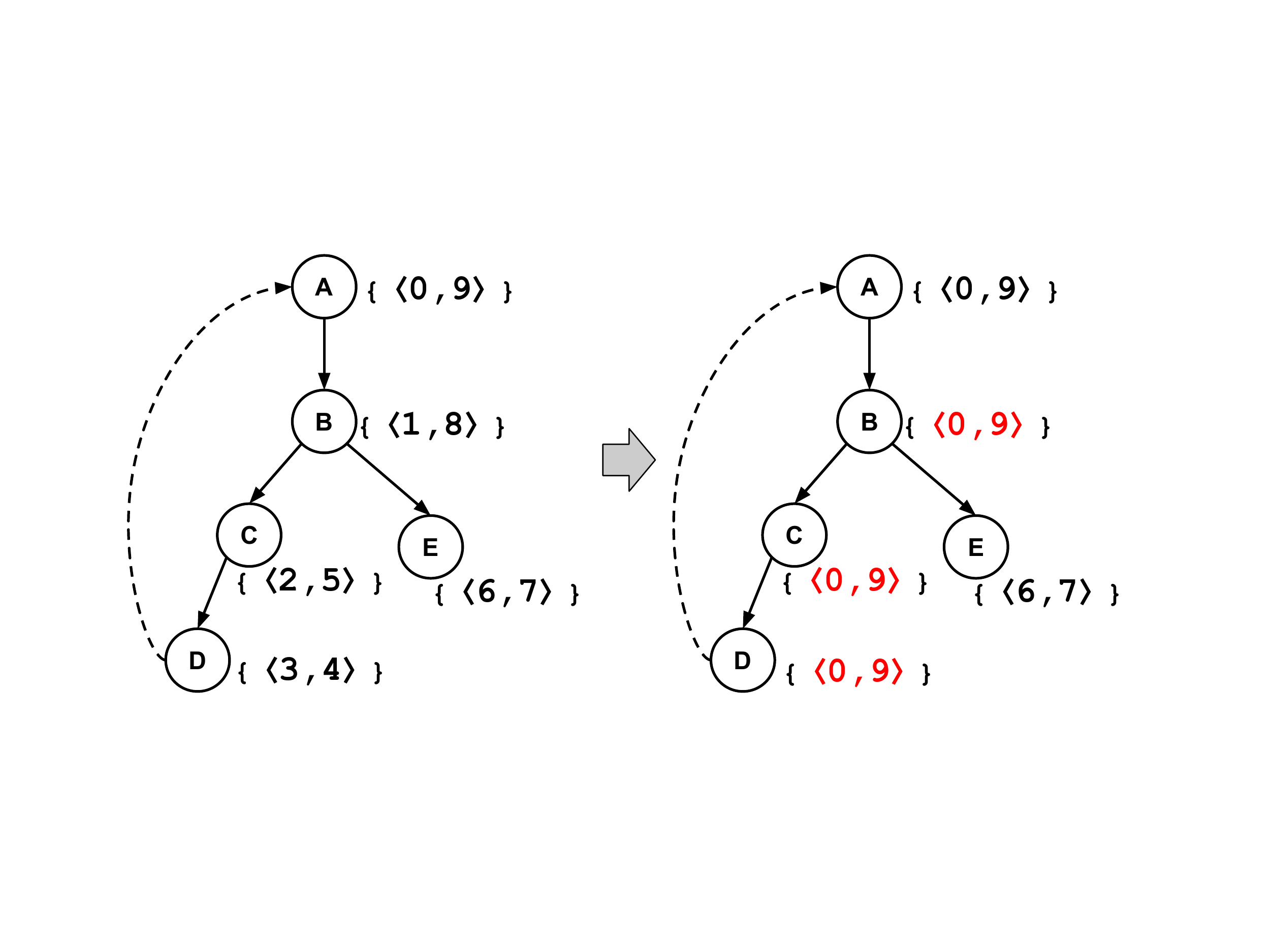}
        }
        \caption[Caption for LOF]{Back Edge}
        \label{fig:dft-interval-back-edge}
    \end{subfigure}
    \caption{Non-tree edge handling}
\end{figure}

Given a cross edge $v_s \to v_d$, $\intervalset{v_d}$ is merged into $\intervalset{v_s}$ and the interval sets of all of its ancestors. Figure~\ref{fig:dft-interval-cross-edge} shows an example of handling cross edge $E \to D$. The interval set for destination vertex $D$, $\{\interval{2, 3}\}$, is merged into the interval sets of $E$, as well as its ancestors $C$ and $A$. This allows us to account for graph reachability of two vertices that passes through $E \to D$. For example, $C$ can reach $D$ because $\intervalset{C} \supseteq \intervalset{D}$.
If $v_s \to v_d$ is a back edge, $\intervalset{v_d}$ will be merged into the interval sets of all vertices in the corresponding Strongly-Connected Component (SCC). The intuition behind this is that every vertex in such SCC belong to a subtree rooted at $v_d$, per the definition of back edge mentioned in Section~\ref{sec:background-dft-intervals}. In other words, $\intervalset{v_d}$ subsumes the interval sets of all of those vertices. Thus, our merging scheme here is able to reflect the mutual connectivity of vertices in a SCC, including the back edge.
Figure~\ref{fig:dft-interval-back-edge} shows an example of handling back edge $D \to A$. The interval set for destination vertex $A$, $\{\interval{0, 9}\}$, is merged into the interval sets of every other vertex in the same SCC. Namely, vertices $B$, $C$, and $D$.
\paragraph{Meet Operator}
The meet operator specifies how value flows are combined from different program paths \emph{e.g.} at a control flow merge point. In our algorithm, we use interval set merge $\cup$ as our meet operator. Per our previous definition of $\cup$, the merged interval set is always a safe approximation for the incoming interval sets and thus ensures the soundness.

\subsection{Inter-procedural Analysis}\label{sec:design-interprocedural}
This Section discusses how to apply our interval-based reachability algorithm across interprocedural constructions \emph{e.g.} call sites.
\coolname{} uses a summary-based interprocedural value-flow algorithm. We summarize the value-flows of each function in a \textit{function value-flow summary} which is propagated to all of its call sites. Since the newly introduced callee summary alters value-flows inside the caller function, this process is repeated until a fixed point is reached.

A function value-flow summary comprises of a set of reachable relationships between function arguments and (outgoing) results. The results of a function are returned values or output (\emph{i.e.} pointer) arguments. The summary describes the mapping between arguments and results which is expressed through argument and result indices. The result index is equal to the index of its counterpart in the result list of the \texttt{dfi.call} operation (see Section~\ref{sec:background-defi-call}). In every \texttt{dfi.call}, the original returned value (if there is any) has index 0, followed by pointer argument type results. We denote $I(p)$ as the argument index of argument $p$. In addition, $O(p)$ is defined as the result index of $p$ if $p$ is part of $PT_f$, a subset of function arguments for $f$ containing all pointer arguments.
The value-flow summary for function $f$ is denoted as $S_f=S^R_f\cup S^P_f$. The sets $S^R_f$ and $S^P_f$ represent mappings from function arguments to return value and to output arguments respectively.
Let $R_f$ and $P_f$ be the set of return values and arguments in $f$, respectively. $S^R_f$ is defined as
\begin{align*}
\{ I(p) \leadsto 0 \mid p \leadsto r, \forall p \in P_f, \forall r \in R_f\}
\end{align*}

\begin{figure}
    \centering
    \begin{subfigure}{.5\columnwidth}
    \resizebox{\columnwidth}{!}{%
       \setlength\tabcolsep{0.9pt}
       \begin{tabular}{|c|c|c|}
          \hline
          \small{Value} & \small{Interval Set} & \small{Summarized Value-Flow} \\
          & & \\
          \hline
          \texttt{\%r} & $\{\langle 0,7\rangle\}$ &
          \multirow{4}{*}{\raisebox{-0.5\height}{\usetikzlibrary{arrows}
\tikzset{>=latex}
\begin{tikzpicture}
    \node[font=\small,rectangle] (A) at (0,0) {\texttt{\%r}};
    \node[font=\small,rectangle] (B) at (0,1) {\texttt{\%a}};
    \node[font=\small,rectangle] (C) at (1,1) {\texttt{\%b}};
    \node[font=\small,rectangle] (D) at (2,1) {\texttt{\%c}};

    \path [->,thick] (B) edge node[font=\small] {} (A);
    \path [->,thick] (D) edge node[font=\small] {} (A);
\end{tikzpicture}}} \\
          \texttt{\%t} & $\{\langle 1,6\rangle\}$ & \\
          \texttt{\%a} & $\{\langle 2,3\rangle\}$ & \\
          \texttt{\%c} & $\{\langle 4,5\rangle\}$ & \\
          \hline
        \end{tabular}
        }
        \caption[Caption for LOF]{Listing~\ref{lst:defi-interval-example2}}
        \label{fig:defi-callee-summaries}
    \end{subfigure}%
    \begin{subfigure}{.5\columnwidth}
    \resizebox{\columnwidth}{!}{%
       \setlength\tabcolsep{0.9pt}
       \begin{tabular}{|c|c|c|}
          \hline
          \small{Value} & \small{Interval Set} & \small{Summarized Value-Flow} \\
          & &\small{(On Call Site)} \\
          \hline
          \texttt{\%t1} & $\{\langle 0,7\rangle\}$ &
          \multirow{4}{*}{\raisebox{-0.5\height}{\usetikzlibrary{arrows}
\tikzset{>=latex}
\begin{tikzpicture}
    \node[font=\small,rectangle] (A) at (0,1) {\texttt{\%k}};
    \node[font=\small,rectangle] (B) at (2,1) {\texttt{\%r}};
    \node[font=\small,rectangle] (C) at (0,0) {\texttt{\%k'}};
    \node[font=\small,rectangle] (D) at (2,0) {\texttt{\%r'}};

    \path [->,thick] (A) edge node[font=\small] {} (C);
    \path [->,thick] (A) edge node[font=\small] {} (D);
    \path [->,thick] (B) edge node[font=\small] {} (C);
    \path [->,thick] (B) edge node[font=\small] {} (D);
\end{tikzpicture}}} \\
          \texttt{\%t0} & $\{\langle 1,4\rangle\}$ & \\
          \texttt{\%k} & $\{\langle 2,3\rangle\}$ & \\
          \texttt{\%r} & $\{\langle 5,6\rangle\}$ & \\
          \hline
        \end{tabular}
        }
        \caption[Caption for LOF]{Listing~\ref{lst:defi-advanced-callee-summaries}, reversed root \texttt{\%t1}}
        \label{fig:defi-advanced-callee-summaries}
    \end{subfigure}
    \caption[Caption for LOF]{Function value-flow summaries}
\end{figure}

Figure~\ref{fig:defi-callee-summaries} shows the value-flow summary of \texttt{f} along with interval sets of relevant values. Since $\intervalset{r} \supseteq \intervalset{a}$ and $\intervalset{r} \supseteq \intervalset{c}$, the value of $S^R_{f} == S_f$ is $\{0 \leadsto 0, 2 \leadsto 0\}$

\begin{minipage}{\columnwidth}
\begin{lstlisting}[
    style=MLIRStyle,
    captionpos=b, caption={A MLIR function with two pointer arguments},
    label={lst:defi-advanced-callee-summaries}]
llvm.func @g(%k: !llvm.ptr<i32>, %r: !llvm.ptr<i32>) {
    %t0 = llvm.load %k : !llvm.ptr<i32>
    %t1 = dfi.store %t0, %r : !llvm.ptr<i32>
}
\end{lstlisting}
\end{minipage}


$S^P_f$ captures the potential value-flows applied on pointer arguments in function $f$.
Specifically, it contains mappings between pointer arguments and corresponding output results on the \texttt{dfi.call} sites.
Since the result of a pointer argument can be difficult to locate in the callee, we use the reachability between pointer arguments and the reversed roots as an approximation.
For arbitrary two pointer arguments $p0$ and $p1$, $S^P_f$ is initialized with $I(p0) \leadsto O(p0)$ and $I(p1) \leadsto O(p1)$. If $p0$ and $p1$ can both reach a reversed root $\tau$, then we can add $I(p0) \leadsto O(p1)$ and $I(p1) \leadsto O(p0)$ into $S^P_f$. Formally speaking, let $\mathrm{T}_f$ be the set of reversed roots in $f$, $S^P_f$ is defined as
\begin{align*}
\{ &I(p) \leadsto O(k), I(k) \leadsto O(p) \mid \exists \tau \in \mathrm{T}_f \emph{ s.t. } \\
   &p \leadsto \tau \land k \leadsto \tau, \forall p, k \in PT_f\}
   \cup \{ I(p) \leadsto O(p) \mid \forall p \in PT_f\}
\end{align*}
Figure~\ref{fig:defi-advanced-callee-summaries} shows an example value-flow summary for function \texttt{g} in Listing~\ref{lst:defi-advanced-callee-summaries}. $k'$ and $r'$ correspond to the results of $k$ and $r$ at a call site, respectively. Since both $k$ and $r$ can reach the reversed root \texttt{\%t1}, $S^P_g$ in this case can be written as
$$
\{I(k) \leadsto O(k), I(k) \leadsto O(r), I(r) \leadsto O(r), I(r) \leadsto O(k)\}
$$
which is equal to $\{0 \leadsto 0, 0 \leadsto 1, 1 \leadsto 1, 1 \leadsto 0\}$

\subsubsection*{Value-Flow Summary Propagation}\label{sec:design-value-flow-propagation}
Next, we propagate the value-flow summary $S_f$ to every call site of $f$. On a high level, $S_f$ provides the missing local value-flow mappings at every call site of $f$. With this information in place, we are able to add new value-flows to the caller functions to improve precision. This process consists of two phases.

In the first phase, for each $v_s \leadsto v_d$ in $S_f$, we perform another round of reversed DFT traversal in every \texttt{caller} function. But this time, instead of using reversed roots designated by the client analysis, we will use the actual parameter values of $v_s$ as revered roots.
\vspace*{-0.5em}
\begin{lstlisting}[
    style=MLIRStyle,
    captionpos=b, caption={A simple MLIR snippet},
    label={lst:defi-callee-summary-propagation}]
llvm.func @f(%v:i32, %p:!llvm.ptr<i32>) -> i32 {
    %t0 = llvm.load %p : !llvm.ptr<i32>
    %t1 = dfi.store %v, %p : !llvm.ptr<i32>
    llvm.return %t0 : i32
}
llvm.func @g(%k:i32, %b:!llvm.ptr<i32>) -> i32 {
    %s0 = llvm.add %k, %k : i32
    %s1,%s2 = dfi.call @f(%s0, %b)
    %s3 = dfi.store %k, %s2 : !llvm.ptr<i32>
    llvm.return %s1 : i32
}
\end{lstlisting}
\vspace*{-0.5em}
In Listing~\ref{lst:defi-callee-summary-propagation}, function \texttt{g} calls function \texttt{f} on line~8. In this snippet, we assume the value-flow summary $S_f$ is $\{0 \leadsto 1, 1\leadsto 0, 1 \leadsto 1\}$.
On line~8, phase 1 performs a reversed DFT traversal using \texttt{\%s0} (\emph{i.e.} the actual parameter \texttt{\%v} in \texttt{f}) and \texttt{\%b} (\emph{i.e.} the actual parameter \texttt{\%p} in \texttt{f}) as reversed roots.

In the second phase, the results from the first phase are propagated in the opposite direction on every call site. In Listing~\ref{lst:defi-callee-summary-propagation} for example, the interval sets for \texttt{\%s0} and \texttt{\%b}, obtained in phase 1, are transitively merged into the interval sets of its parent and ancestor vertices (\emph{e.g.} \texttt{\%s3}) in the augmented DFT, until reaching the original reversed roots.

As mentioned earlier, propagating callee value-flow summaries in the caller function might change its value-flow summary as well. \coolname{} uses a worklist-based algorithm, shown in Algorithm~\ref{algo:defi-interprocedual-worklist}, to recursively propagate the changed summaries to its callers until reaching a fixed point.
\begin{algorithm}[H]
\caption{Interprocedural Worklist Algorithm}
\label{algo:defi-interprocedual-worklist}
\scriptsize
\begin{algorithmic}[1]
\State $Worklist = $ all functions
\While{$Worklist \neq \emptyset$}
    \State $f := Worklist.pop()$
    \State $S_f := GetVFSummary(f)$
    \If{$S_f $ has changed}
        \ForAll{$c \in callers(f)$}
            \State $PropagateSummary(S_f, c)$
            \If{$c \notin Worklist$}
                \State $Worklist.push(c)$
            \EndIf
        \EndFor
    \EndIf
\EndWhile
\end{algorithmic}
\end{algorithm}
In Algorithm~\ref{algo:defi-interprocedual-worklist}, $GetVFSummary(f)$ returns $S_f$ for function $f$; $PropagateSummary(S_f, c)$ performs the two-phase propagation introduced earlier with callee value-flow summary $S_f$ in caller function $c$.

\subsubsection*{Reachable Functions Summary}\label{sec:design-interprocedural-function-summary}
We are now able to calculate interprocedural value-flows with flow and context-sensitivity of two values $v_a$ and $v_b$ in the \textit{same} function by determining their reachability via interval sets \emph{i.e.} $\intervalset{v_b} \supseteq \intervalset{v_a}$. In this part, we generalize this ability for $v_a$ and $v_b$ that reside in arbitrary functions.

In the previous Section, the value-flow summary of callee function was propagated to the caller, in order to add context-sensitive value-flows into the caller function. This process is now augmented to additionally propagate a \textit{reachable function summary}. A reachable function summary $\Psi(\varepsilon)$ provides a mapping from an endpoint $\varepsilon$ to a set of \textit{transitively-reachable} endpoints. An endpoint $\varepsilon^i_f$ represents the $i$-th argument of function $f$.
\begin{lstlisting}[
    style=MLIRStyle,
    captionpos=b, caption={MLIR functions with interprocedural calls},
    label={lst:defi-reachable-function-summary}]
llvm.func @f(%a0: i32, %a1: i32) {
    dfi.call @g(%a0)
}
llvm.func @g(%a0:i32) {
    dfi.call @f(10, %a0)
    dfi.call @k(23, 75, %a0)
}
llvm.func @k(%a0:i32, %a1:i32, %a2:i32)
\end{lstlisting}
Take Listing~\ref{lst:defi-reachable-function-summary} as an example, the reachable function summary for $\varepsilon^0_f$ (\emph{i.e.} the first argument of function \texttt{f}) is
$$
\Psi(\varepsilon^0_f) = \{\varepsilon^0_g, \varepsilon^1_f, \varepsilon^2_k\}
$$
Because the first argument of \texttt{f} will be passed to the first argument of \texttt{g}, which further passes it to the second and third argument of \texttt{f} and \texttt{k}, respectively.

To calculate $\Psi$, we use the same infrastructure outlined in Algorithm~\ref{algo:defi-interprocedual-worklist}. Basically, for a given function $f$ with $n$ arguments, $\Psi(\varepsilon^{0...n-1}_f)$ is repeatedly propagated to all of its callers and merged with their summaries until reaching a fixed point, namely, none of the reachable function summaries changed.
With reachable function summaries, we are able to perform a two-stage process to answer the value-flow query $v_a \leadsto v_b$, where $v_a$ and $v_b$ are in different functions $f$ and $g$, respectively. First, we calculate a set
$$
\{\varepsilon^i_e \mid \forall e \in callees(f) \emph{ s.t. } v_a \leadsto \varepsilon^i_e\}
$$
Then, we check if any of the $\Psi(\varepsilon^i_e)$ contains an endpoint of $g$, $\varepsilon^j_g$, for arbitrary argument index $j$. If not, that means function $f$ cannot even reach function $g$. Finally, in the second stage, we can conclude that $v_a$ can reach $v_b$ only if endpoint $\varepsilon^j_g$ can reach $v_b$. Namely, $\varepsilon^j_g \leadsto v_b \implies v_a \leadsto v_b$.

\section{Evaluation}\label{sec:eval}
In this Section, we discuss two research questions:
\begin{itemize}
    \item \textbf{RQ1:} How well does \coolname{} scale on large codebases and how does it compare against other state-of-the-art value-flow analysis tools? (\S~\ref{sec:eval-perf} and \S~\ref{sec:eval-perf-comparison})
    \item \textbf{RQ2:} The efficiency of answering graph reachability in \coolname{} is strongly related to the size of the interval set associated with each vertex. What is the size distribution of interval sets across all vertices? (\S~\ref{sec:eval-interval-set-size})
\end{itemize}
We selected four open-source codebases from different domains and of different sizes as our analysis targets: Lighttpd, SQLite 3, OpenSSL, and FFmpeg. Table~\ref{tab:benchmarks-description} details the selected target programs,
lines-of-code (LOC) are listed in textual as well as LLVM 14 IR code, as \coolname{} operates on LLVM IR input.
\begin{table}[]
    \centering
    \resizebox{\columnwidth}{!}{%
    \begin{tabular}{|c|c|c|c|c|c|}
    \hline
    Target  & Version & \# LOC & \# LOC (LLVM IR) & \# of Functions & Description             \\ \hline
    FFmpeg   & 4.2     & 1.2M   & 4.6M             & 17802           & A/V encoder and decoder \\ \hline
    OpenSSL  & 3.0.0   & 532K   & 1.1M             & 13490           & Crypto and TLS          \\ \hline
    SQLite 3 & 3.36.0  & 166K   & 376K             & 1103            & SQL database            \\ \hline
    Lighttpd & 1.4.60  & 97K    & 216K             & 1258            & Web server              \\ \hline
    \end{tabular}%
    }
    \caption{Description of target codebases}
    \label{tab:benchmarks-description}
\end{table}
The targets are built using Link-Time Optimization (LTO) to generate a monolithic executable without dynamic dependencies.
All experiments are performed on a commodity desktop running Ubuntu 20.04 LTS with a 6-core, 12-thread Intel i7-8700K CPU and 32GB of RAM.

\subsection{Performance of Different Client Analyses}\label{sec:eval-perf}
We measured the performance of \coolname{} with two different client analyses: taint analysis and read-only argument analysis, both analyses are flow and context-sensitive.

The taint analysis is designed to taint address-taken variables. If the address-taken variable is tainted, all related pointer variables comprising of variables pointing to the beginning of the object as well as pointers derived from an offset, are tainted as well; any value loaded from a tainted variable is also considered tainted.
Our implementation traverses the following operations: memory load, \texttt{dfi.store}, and \texttt{getelementptr} which creates a new pointer by applying an offset to a base pointer.

Read-only argument analysis detects whether a pointer argument is modified in the callee function. A pointer argument $p$ of callee function $f$ is considered modified (\emph{i.e.} not read-only) on a \texttt{dfi.call} call site if any of the argument values, other than $p$, can reach the corresponding result of $p$ (\emph{i.e.} result at index $O(p)$) on that call site.
Our implementation traverses pointer values within $f$ through their def-use chains. For all indirect memory stores on the path, we also traverse the stored value and its def-use chains.

\begin{figure}
\pgfplotstableread[
   col sep=comma,
]{./perbreak.csv}\sccattable
\pgfplotsset{
  every tick label/.append style={font=\tiny},
}

\begin{minipage}{\columnwidth}
\centering
\begin{tikzpicture}
    \begin{axis}[%
    hide axis,
    xbar,
    xmin=10, xmax=50, ymin=0, ymax=0.4,
    legend style={legend cell align=left, draw=none},
    width=0.23*\columnwidth,
    axis lines=none,
    ]
    \addlegendimage{black, fill=lightgray2, postaction={pattern=north east lines}}
    \addlegendentry{\tiny{Parsing}};
    \end{axis}
\end{tikzpicture}
\begin{tikzpicture}
    \begin{axis}[%
    hide axis,
	 xbar,
    xmin=10, xmax=50, ymin=0, ymax=0.4,
    legend style={legend cell align=left, draw=none},
    width=0.23*\columnwidth,
    axis lines=none,
    ]
    \addlegendimage{black, fill=lightgray1, postaction={pattern=dots}}
    \addlegendentry{\tiny{Preprocess}};
    \end{axis}
\end{tikzpicture}
\begin{tikzpicture}
    \begin{axis}[%
    hide axis,
	 xbar,
    xmin=10, xmax=50, ymin=0, ymax=0.4,
    legend style={legend cell align=left, draw=none},
    width=0.23*\columnwidth,
    axis lines=none,
    ]
    \addlegendimage{black, fill=darkred}
    \addlegendentry{\tiny{Analysis}};
    \end{axis}
\end{tikzpicture}
\begin{tikzpicture}
    \begin{axis}[%
    hide axis,
	 xbar,
    xmin=10, xmax=50, ymin=0, ymax=0.4,
    legend style={legend cell align=left, draw=none},
    width=0.23*\columnwidth,
    axis lines=none,
    ]
    \addlegendimage{black, fill=lightgray2, postaction={pattern=crosshatch}}
    \addlegendentry{\tiny{Verification}};
    \end{axis}
\end{tikzpicture}
\begin{tikzpicture}
    \begin{axis}[%
    hide axis,
	 xbar,
    xmin=10, xmax=50, ymin=0, ymax=0.4,
    legend style={legend cell align=left, draw=none},
    width=0.23*\columnwidth,
    axis lines=none,
    ]
    \addlegendimage{black, fill=lightgray1, postaction={pattern=north west lines, pattern color=black}}
    \addlegendentry{\tiny{Output}};
    \end{axis}
\end{tikzpicture}
\end{minipage}\\
\begin{tikzpicture}
   \begin{axis}[
         xbar stacked,
         yticklabels from table={\sccattable}{target},
         ytick=data,
         xmax=1.0,
         xmin=0.0,
         height=3cm,
         bar width=0.3cm,
         width=\columnwidth,
      ]
      \addplot[black, fill=lightgray2, postaction={pattern=north east lines}] table [
         col sep=comma,
         x expr=\thisrow{Parsing}/\thisrow{total},
         y expr=\coordindex,
      ]{\sccattable};
      \addplot[black, fill=lightgray1, postaction={pattern=dots}] table [
         col sep=comma,
         x expr=\thisrow{Preprocess}/\thisrow{total},
         y expr=\coordindex,
      ]{\sccattable};
      \addplot[black, fill=darkred] table [
         col sep=comma,
         x expr=\thisrow{Analysis}/\thisrow{total},
         y expr=\coordindex,
      ]{\sccattable};
      \addplot[black, fill=lightgray2, postaction={pattern=crosshatch}] table [
         col sep=comma,
         x expr=\thisrow{Verification}/\thisrow{total},
         y expr=\coordindex,
      ]{\sccattable};
      \addplot[black, fill=lightgray1, postaction={pattern=north west lines, pattern color=black}] table [
         col sep=comma,
         x expr=\thisrow{Output}/\thisrow{total},
         y expr=\coordindex,
      ]{\sccattable};
   \end{axis}
\end{tikzpicture}
  \caption{Timing breakdown of read-only argument analysis}
  \label{fig:eval-perf-breakdown}
\end{figure}
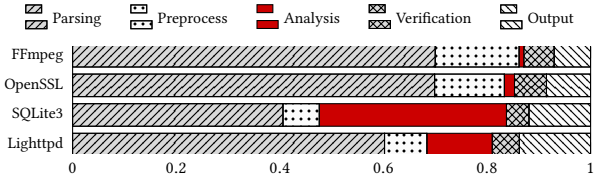

Table~\ref{tab:defi-performance} lists the results of running our two client applications against the targets listed in Table~\ref{tab:benchmarks-description}.
The runtime measurement is split into two categories, \textit{Total time} and \textit{Analysis time}. The total time accounts for the wall-clock time spent on the entire process, while the analysis time only measures the time spent in the main analysis (see Figure~\ref{fig:eval-perf-breakdown} for an overview of processing steps before and after the main analysis).
The memory consumption measures the maximum amount of physical memory allocated during the evaluation of each target, which is roughly equal to resident memory.
The number of visited edges and vertices is listed in the \textit{\#V-Edge} and \textit{\#V-Vertex} columns, respectively.
Our evaluation shows that read-only argument analysis visited roughly 20\% $\sim$ 60\% more vertices and edges compared to taint analysis on every target, except for FFmpeg, where the number of visited edges increased by 160\% for taint analysis.
All experiments finished within \textit{10 minutes} using no more than \textit{4.5 GB} of physical memory. In the next two paragraphs, we will proceed to breakdown the performance and discuss the scalability w.r.t different codebase sizes.
\begin{table*}[]
    \centering
    \resizebox{\textwidth}{!}{%
    \begin{tabular}{|c|ccccc|ccccc|}
    \hline
     &
      \multicolumn{5}{c|}{Taint Analysis} &
      \multicolumn{5}{c|}{Read-Only Argument Analysis} \\ \hline
    Target &
      \multicolumn{1}{c|}{\#V-Edge} &
      \multicolumn{1}{c|}{\#V-Vertex} &
      \multicolumn{1}{c|}{Total time} &
      \multicolumn{1}{c|}{Analysis time} &
      Max resident memory &
      \multicolumn{1}{c|}{\#V-Edge} &
      \multicolumn{1}{c|}{\#V-Vertex} &
      \multicolumn{1}{c|}{Total time} &
      \multicolumn{1}{c|}{Analysis time} &
      Max resident memory \\ \hline
    FFmpeg &
      \multicolumn{1}{c|}{730.0K} &
      \multicolumn{1}{c|}{1.53M} &
      \multicolumn{1}{c|}{497.66s} &
      \multicolumn{1}{c|}{2.78s} &
      3.19 GB &
      \multicolumn{1}{c|}{1.9M} &
      \multicolumn{1}{c|}{2.6M} &
      \multicolumn{1}{c|}{534.63s} &
      \multicolumn{1}{c|}{5.42s} &
      4.08 GB \\ \hline
    OpenSSL &
      \multicolumn{1}{c|}{250.9K} &
      \multicolumn{1}{c|}{549.1K} &
      \multicolumn{1}{c|}{99.99s} &
      \multicolumn{1}{c|}{1.15s} &
      876.60 MB &
      \multicolumn{1}{c|}{398.4K} &
      \multicolumn{1}{c|}{677.3K} &
      \multicolumn{1}{c|}{101.47s} &
      \multicolumn{1}{c|}{2.05s} &
      1.09 GB \\ \hline
    SQLite 3 &
      \multicolumn{1}{c|}{81.4K} &
      \multicolumn{1}{c|}{156.7K} &
      \multicolumn{1}{c|}{3.47s} &
      \multicolumn{1}{c|}{564.0ms} &
      337.55 MB &
      \multicolumn{1}{c|}{128.1K} &
      \multicolumn{1}{c|}{202.3K} &
      \multicolumn{1}{c|}{4.72s} &
      \multicolumn{1}{c|}{1.71s} &
      424.19 MB \\ \hline
    Lighttpd &
      \multicolumn{1}{c|}{27.2K} &
      \multicolumn{1}{c|}{62.0K} &
      \multicolumn{1}{c|}{1.05s} &
      \multicolumn{1}{c|}{104.2ms} &
      168.10 MB &
      \multicolumn{1}{c|}{42.2K} &
      \multicolumn{1}{c|}{75.6K} &
      \multicolumn{1}{c|}{1.09s} &
      \multicolumn{1}{c|}{142.3ms} &
      189.07 MB \\ \hline
    \end{tabular}%
    }
    \caption{Performance and memory consumption of \coolname{}}
    \label{tab:defi-performance}
\end{table*}
\vspace*{-0.7em}
\subsubsection*{Performance Breakdown}
\coolname{} is built using the standard MLIR framework. We use the built-in MLIR parser to parse our input and process it with a MLIR pass pipeline that runs the preprocessor and analysis engine.
In this pipeline, we found that some built-in components that are not part of the \coolname{} codebase comprise the majority of execution time. The total pipeline execution time can be broken down into five major components which are listed in Figure~\ref{fig:eval-perf-breakdown}.
We evaluate the ratio of processing time spent in each component for read-only argument analysis across all targets. Among them, the MLIR parser always accounts for most of the runtime. Other large contributors are the verifier, which verifies MLIR code after each pass, and the output stage, these nonfactors account for 55\% $\sim$ 85\% of the total execution time.
The preprocessor is part of \coolname{} (see Section~\ref{sec:preprocessing}), though we argue that its runtime is a single-shot cost that will be amortized by the number of MLIR passes in the pipeline.
For the scalability evaluation we will therefore focus on the runtime and memory requirements of the main analysis, which constitutes the core contribution of this paper.
\begin{figure}
  \setlength\tabcolsep{0pt}
\pgfplotstableread[
    col sep=comma,
]{./scale.csv}\sccattable
\centering
\begin{minipage}{\linewidth}
   \centering
    \begin{tikzpicture}
        \begin{axis}[width=0.24*\columnwidth, hide axis, xmin=0, xmax=0, ymin=0,ymax=0, legend style={font=\tiny,legend cell align=left, draw=none}]
            \addlegendimage{thick, densely dotted, mark=triangle*, mark size=1.2pt, mark options=solid}
            \addlegendentry{Visited Edges};
        \end{axis}
    \end{tikzpicture}
    \begin{tikzpicture}
        \begin{axis}[width=0.24*\columnwidth, hide axis, xmin=0, xmax=0, ymin=0,ymax=0, legend style={font=\tiny,legend cell align=left, draw=none}]
            \addlegendimage{thick, densely dashed, mark=diamond*,mark size=1.2pt, mark options=solid}
            \addlegendentry{Visited Vertices};
        \end{axis}
    \end{tikzpicture}
    \begin{tikzpicture}
        \begin{axis}[width=0.24*\columnwidth, hide axis, xmin=0, xmax=0, ymin=0,ymax=0, legend style={font=\tiny,legend cell align=left, draw=none}]
           \addlegendimage{ultra thick, darkred, mark=o,mark size=1.2pt, mark options={solid, fill=darkred}}
            \addlegendentry{Analysis Time, Max Memory};
        \end{axis}
    \end{tikzpicture}
\end{minipage}
    \begin{tabular}{l c c r}
        & \tiny{Taint Analysis} & \tiny{RO Argument Analysis} & \\
        \multirow{2}{*}[-0.5ex]{\rotatebox[origin=c]{90}{\tiny{\# Visited Edges, Vertices $\times 10^6$}}}
        &
        \raisebox{-0.5\height}{
        \begin{tikzpicture}
        \begin{axis}[
                xticklabels={,,},
                xtick=data,
                y tick label style={font=\tiny},
                width=4.5cm,
                height=3.8cm,
                ytick pos=right,
                xtick pos=left,
                yticklabel pos=right,
                ytick = {0, 1.0, 2.0, 3.0, 4.0},
                ymajorgrids=true,
                ymin=0,
                ymax=4.0,
                xmin=0,
                xmax=3,
            ]
            \addplot [ultra thick, darkred, mark=o, mark size=1.2pt, mark options={solid, fill=darkred}] table [
                col sep=comma,
                x expr=\coordindex,
                y=TT,
            ]{\sccattable};
        \end{axis}
        \begin{axis}[
                xticklabels={,,},
                xtick=data,
                ytick pos=left,
                xtick pos=left,
                y tick label style={font=\tiny},
                ytick = {0, 0.5, 1.0, 1.5, 2.0},
                width=4.5cm,
                height=3.8cm,
                ymin=0,
                ymax=2.0,
                xmin=0,
                xmax=3,
            ]
            \addplot [thick, densely dotted, mark=triangle*, mark size=1.2pt, mark options=solid] table [
                col sep=comma,
                x expr=\coordindex,
                y=VET,
            ]{\sccattable};
            \addplot [thick, densely dashed, mark=diamond*, mark size=1.2pt, mark options=solid] table [
                col sep=comma,
                x expr=\coordindex,
                y=VVT,
            ]{\sccattable};
        \end{axis}
    \end{tikzpicture}
    }
    &
    \raisebox{-0.5\height}{
    \begin{tikzpicture}
        \begin{axis}[
                xticklabels={,,},
                xtick=data,
                y tick label style={font=\tiny},
                width=4.5cm,
                height=3.8cm,
                ytick pos=right,
                xtick pos=left,
                yticklabel pos=right,
                ymajorgrids=true,
                ytick = {0, 2.0, 4.0, 6.0},
                ymin=0,
                ymax=6.0,
                xmin=0,
                xmax=3,
            ]
            \addplot [ultra thick, darkred, mark=o, mark size=1.2pt, mark options={solid, fill=darkred}] table [
                col sep=comma,
                x expr=\coordindex,
                y=TA,
            ]{\sccattable};
        \end{axis}
        \begin{axis}[
                xticklabels={,,},
                xtick=data,
                ytick pos=left,
                xtick pos=left,
                y tick label style={font=\tiny},
                ytick = {0, 1.0, 2.0, 3.0},
                width=4.5cm,
                height=3.8cm,
                ymin=0,
                ymax=3.0,
                xmin=0,
                xmax=3,
            ]
            \addplot [thick, densely dotted, mark=triangle*, mark size=1.2pt, mark options=solid] table [
                col sep=comma,
                x expr=\coordindex,
                y=VEA,
            ]{\sccattable};
            \addplot [thick, densely dashed, mark=diamond*, mark size=1.2pt, mark options=solid] table [
                col sep=comma,
                x expr=\coordindex,
                y=VVA,
            ]{\sccattable};
        \end{axis}
    \end{tikzpicture}
    }
    &
    \raisebox{0.0\height}{
    \rotatebox[origin=c]{90}{\tiny{Time (sec)}}
    }
    \\
    &
    \raisebox{-0.5\height}{
    \begin{tikzpicture}
        \begin{axis}[
                xticklabels from table={\sccattable}{target},
                xtick=data,
                x tick label style={rotate=90,font=\tiny},
                y tick label style={font=\scriptsize},
                ylabel near ticks,
                xlabel near ticks,
                ytick = {0, 1.0, 2.0, 3.0, 4.0},
                width=4.5cm,
                height=3.8cm,
                ymajorgrids=true,
                ytick pos=right,
                xtick pos=left,
                ylabel near ticks, yticklabel pos=right,
                ymin=0,
                ymax=4.0,
                xmin=0,
                xmax=3,
            ]
            \addplot [ultra thick, darkred, mark=o, mark size=1.2pt, mark options={solid, fill=darkred}] table [
                col sep=comma,
                x expr=\coordindex,
                y=MT,
            ]{\sccattable};
        \end{axis}
        \begin{axis}[
                xticklabels={,,},
                xtick=data,
                ytick pos=left,
                xtick pos=left,
                y tick label style={font=\tiny},
                ytick = {0, 0.5, 1.0, 1.5, 2.0},
                width=4.5cm,
                height=3.8cm,
                ymax=2.0,
                ymin=0,
                xmin=0,
                xmax=3,
            ]
            \addplot [thick, densely dotted, mark=triangle*, mark size=1.2pt, mark options=solid] table [
                col sep=comma,
                x expr=\coordindex,
                y=VET,
            ]{\sccattable};
            \addplot [thick, densely dashed, mark=diamond*, mark size=1.2pt, mark options=solid] table [
                col sep=comma,
                x expr=\coordindex,
                y=VVT,
            ]{\sccattable};
        \end{axis}

    \end{tikzpicture}
}
        &
   \raisebox{-0.5\height}{
    \begin{tikzpicture}
        \begin{axis}[
                xticklabels from table={\sccattable}{target},
                xtick=data,
                x tick label style={rotate=90,font=\tiny},
                y tick label style={font=\scriptsize},
                xlabel near ticks,
                width=4.5cm,
                height=3.8cm,
                ytick = {0, 2.0, 4.0, 6.0},
                ymajorgrids=true,
                ytick pos=right,
                xtick pos=left,
                ylabel near ticks, yticklabel pos=right,
                ymin=0,
                ymax=6.0,
                xmin=0,
                xmax=3,
            ]
            \addplot [ultra thick, darkred, mark=o, mark size=1.2pt, mark options={solid, fill=darkred}] table [
                col sep=comma,
                x expr=\coordindex,
                y=MA,
            ]{\sccattable};
        \end{axis}
        \begin{axis}[
                xticklabels={,,},
                xtick=data,
                ytick pos=left,
                xtick pos=left,
                x tick label style={rotate=45,font=\tiny},
                y tick label style={font=\scriptsize},
                ytick = {0, 1.0, 2.0, 3.0},
                width=4.5cm,
                height=3.8cm,
                ymin=0,
                ymax=3.0,
                xmin=0,
                xmax=3,
            ]
            \addplot [thick, densely dotted, mark=triangle*, mark size=1.2pt, mark options=solid] table [
                col sep=comma,
                x expr=\coordindex,
                y=VEA,
            ]{\sccattable};
            \addplot [thick, densely dashed, mark=diamond*, mark size=1.2pt, mark options=solid] table [
                col sep=comma,
                x expr=\coordindex,
                y=VVA,
            ]{\sccattable};
        \end{axis}

    \end{tikzpicture}
}
    &
    \raisebox{0.5\height}{
    \rotatebox[origin=c]{90}{\tiny{Memory (GB)}}
 }
    \\
    \end{tabular}
  \caption{Scalability of time and memory consumption}
  \label{fig:eval-scalability-chart}
\end{figure}
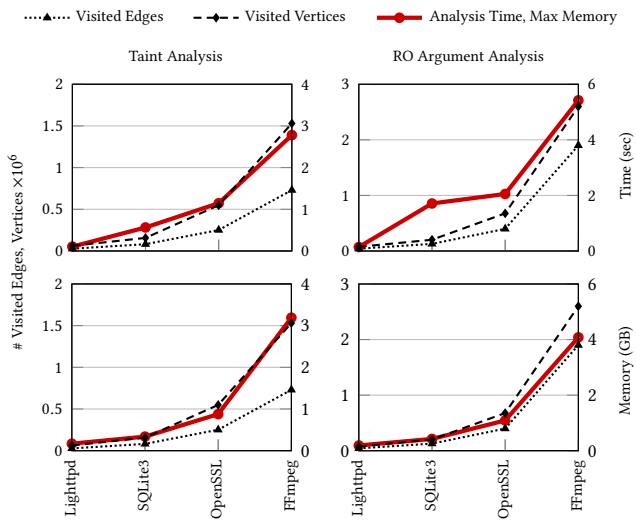
\vspace*{-0.7em}
\subsubsection*{Scalability of the Analysis}
To evaluate the scalability of \coolname{}, we measure the increase in analysis time and memory consumption against increases of the target code size as represented by the number of edges and vertices in \coolname{}'s program graph.
The results of the scalability evaluation are shown in Figure~\ref{fig:eval-scalability-chart}. On the left hand side of the figure we measure the runtime (top) and memory consumption (bottom) of \coolname{} based taint analysis across all targets. The results demonstrate that the runtime as well as the memory consumption of \coolname{} increases by a factor that is bounded by the increase in the number of vertices and edges respectively.
The same experiment is repeated for read-only argument analysis, shown in the right half of the same figure, for which we measure a comparable result.


We notice two exceptions \emph{i.e.} analysis time of both analyses on SQLite 3. We found that, in processing the SQLite 3 codebase, significantly more time was spent in the interprocedural worklist algorithm (Algorithm~\ref{algo:defi-interprocedual-worklist}). Specifically, the propagation of callee function summaries to call sites in every caller functions.
Further investigation reveals two important factors contributing to this problem. First, SQLite 3 has a much \textit{denser} call graph, in which each callee function is called by 3x $\sim$ 4x more call sites on average, compared to call graphs in other targets. Second, Table~\ref{tab:benchmarks-description} shows that functions in SQLite 3 contain more code on average, primarily because the codebase has fewer functions compared to other projects of comparable size.
The fact that these two factors multiply together (\emph{i.e.} higher number of propagations to many large-size caller functions) induces a higher performance overhead in Algorithm~\ref{algo:defi-interprocedual-worklist}.
However, we argue that the source code structure of SQLite 3, more specifically, their interprocedural function call structure, is relatively unusual.
\begin{table}[]
  \centering
  \resizebox{\columnwidth}{!}{%
  \begin{tabular}{|c|cc|cc|cc|}
  \hline
          & \multicolumn{2}{c|}{\coolname{}}            & \multicolumn{2}{c|}{Phasar~\cite{schubertPhASARInterproceduralStatic2019}} & \multicolumn{2}{c|}{SVF~\cite{sui2016svf}}       \\ \hline
  Target  & \multicolumn{1}{c|}{Total time} & Max memory & \multicolumn{1}{c|}{Total time} & Max memory & \multicolumn{1}{c|}{Total time} & Max memory \\ \hline
  FFmpeg  & \multicolumn{1}{c|}{497.66s} & 3.19 GB   & \multicolumn{1}{c|}{-}  & - & \multicolumn{1}{c|}{N/A} & OOM \\ \hline
  OpenSSL & \multicolumn{1}{c|}{99.99s}  & 876.60 MB & \multicolumn{1}{c|}{-}  & - & \multicolumn{1}{c|}{N/A} & OOM \\ \hline
  SQLite 3 & \multicolumn{1}{c|}{3.47s}      & 337.55 MB  & \multicolumn{1}{c|}{N/A}        & OOM        & \multicolumn{1}{c|}{192.04s}    & 6.31 GB    \\ \hline
  Lighttpd & \multicolumn{1}{c|}{1.05s}      & 168.10 MB  & \multicolumn{1}{c|}{24.97s}     & 471.28 MB  & \multicolumn{1}{c|}{67.71s}     & 1.01 GB    \\ \hline
  \end{tabular}%
  }
  \caption{Comparison of taint analysis performance and maximum resident memory consumption of \coolname{} with Phasar and SVF. OOM means out-of-memory; "-" means runtime error.}
  \label{tab:defi-performance-comparison}
\end{table}
\vspace*{-0.5em}
\subsection{Comparison with Other Tools}\label{sec:eval-perf-comparison}
Table~\ref{tab:defi-performance-comparison} compares the performance of \coolname{} against SVF and Phasar.
Phasar computes value-flow information using the IFDS algorithm; SVF utilizes SSA def-use chains and incorporates address taken variables based on precomputed points-to information.
Note that, for this evaluation, we measure the total time rather than analysis time for \coolname{}, in order to provide a fair comparison.
\coolname{} and Phasar are evaluated by performing taint analysis.
For SVF we conservatively estimate a lower bound for its runtime and memory consumption. SVF builds a special Static Value-Flow Graph (SVFG), before running the actual analysis. We found that SVF either runs out of memory or runs slower than \coolname{} during the graph construction phase. Thus, we only include performance of the SVF graph construction phase.
As depicted in Table~\ref{tab:defi-performance-comparison} Phasar was unable to perform the analysis on FFmpeg and OpenSSL due to segmentation faults; it ran out of memory on SQLite 3; on Lighttpd, \coolname{} ran 23x faster with 2.8x less memory than Phasar.
SVF ran out of memory on FFmpeg and OpenSSL; on SQLite 3 and Lighttpd, \coolname{} ran 55x $\sim$ 64x faster and consumes 6x $\sim$ 19x less memory than SVF.
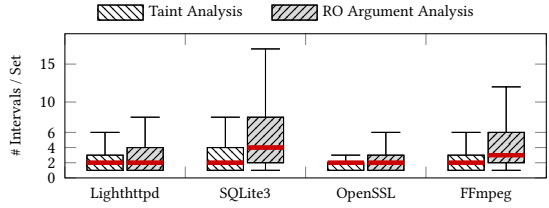
\begin{figure}
  \centering
\makeatletter
\tikzset{nomorepostaction/.code=\let\tikz@postactions\pgfutil@empty}
\makeatother

\begin{minipage}{\columnwidth}
\centering
\begin{tikzpicture}
    \begin{axis}[%
    area legend,
    xmin=10, xmax=50, ymin=0, ymax=0.4,
    hide axis,
    legend style={legend cell align=left, draw=none},
    width=0.23*\columnwidth,
    axis lines=none,
    ]
    \addlegendimage{
    black,
    fill=lightgray1,
    postaction={pattern=north west lines, pattern color=black},
    }
    \addlegendentry{\tiny{Taint Analysis}};
    \end{axis}
\end{tikzpicture}
\begin{tikzpicture}
    \begin{axis}[%
    area legend,
    xmin=10, xmax=50, ymin=0, ymax=0.4,
    hide axis,
    legend style={legend cell align=left, draw=none},
    width=0.23*\columnwidth,
    axis lines=none,
    ]
    \addlegendimage{
    black,
    fill=lightgray2,
    postaction={pattern=north east lines, pattern color=black},
    }
    \addlegendentry{\tiny{RO Argument Analysis}};
    \end{axis}
\end{tikzpicture}
\end{minipage}\\

\begin{tikzpicture}
  \begin{axis}
    [
    boxplot/draw direction=y,
    ylabel={\tiny{\# Intervals / Set}},
    ylabel near ticks,
    xlabel near ticks,
    height=3.5cm,
    width=0.8\columnwidth,
    ymin=0,
    ymax=19,
    xmin=0,
    xmax=4,
    boxplot={
       %
          draw position={1/3 + floor(\plotnumofactualtype/2) + 1/3*mod(\plotnumofactualtype,2)},
       %
          box extend=0.3,
    },
   x=1.6cm,
   xtick={0,1,2,3,4},
   ytick={0,2,4,6,10,15},
   x tick label as interval,
   xticklabels={%
      {\tiny{Lighthttpd}},%
      {\tiny{SQLite3}},%
      {\tiny{OpenSSL}},%
      {\tiny{FFmpeg}},%
   },
   x tick label style={
        font=\tiny,
        text width=2.0cm,
        align=center,
    },
   y tick label style={
      font=\tiny,
   },
   ]
    \addplot+[
    solid,
    black,
    fill=lightgray1,
    every path/.style={postaction={nomorepostaction,pattern=north west lines, pattern color=black}},
    boxplot prepared={
      median=2,
      upper quartile=3,
      lower quartile=1,
      upper whisker=6,
      lower whisker=1,
      every median/.style={darkred,ultra thick},
    },
    ] coordinates {};
    \addplot+[
    solid,
    black,
    fill=lightgray2,
    every path/.style={postaction={nomorepostaction,pattern=north east lines, pattern color=black}},
    boxplot prepared={
      median=2,
      upper quartile=4,
      lower quartile=1,
      upper whisker=8,
      lower whisker=1,
      every median/.style={darkred,ultra thick},
    },
    ] coordinates {};
    \addplot+[
    solid,
    black,
    fill=lightgray1,
    every path/.style={postaction={nomorepostaction,pattern=north west lines, pattern color=black}},
    boxplot prepared={
      median=2,
      upper quartile=4,
      lower quartile=1,
      upper whisker=1,
      lower whisker=8,
      every median/.style={darkred,ultra thick},
    },
    ] coordinates {};
    \addplot+[
    solid,
    black,
    fill=lightgray2,
    every path/.style={postaction={nomorepostaction,pattern=north east lines, pattern color=black}},
    boxplot prepared={
      median=4,
      upper quartile=8,
      lower quartile=2,
      upper whisker=17,
      lower whisker=1,
      every median/.style={darkred,ultra thick},
    },
    ] coordinates {};

    \addplot+[
    solid,
    black,
    fill=lightgray1,
    every path/.style={postaction={nomorepostaction,pattern=north west lines, pattern color=black}},
    boxplot prepared={
      median=2,
      upper quartile=2,
      lower quartile=1,
      upper whisker=3,
      lower whisker=1,
      every median/.style={darkred,ultra thick},
    },
    ] coordinates {};
    \addplot+[
    solid,
    black,
    fill=lightgray2,
    every path/.style={postaction={nomorepostaction,pattern=north east lines, pattern color=black}},
    boxplot prepared={
      median=2,
      upper quartile=3,
      lower quartile=1,
      upper whisker=6,
      lower whisker=1,
      every median/.style={darkred,ultra thick},
    },
    ] coordinates {};

    \addplot+[
    solid,
    black,
    fill=lightgray1,
    every path/.style={postaction={nomorepostaction,pattern=north west lines, pattern color=black}},
    boxplot prepared={
      median=2,
      upper quartile=3,
      lower quartile=1,
      upper whisker=6,
      lower whisker=1,
      every median/.style={darkred,ultra thick},
    },
    ] coordinates {};
    \addplot+[
    solid,
    black,
    fill=lightgray2,
    every path/.style={postaction={nomorepostaction,pattern=north east lines, pattern color=black}},
    boxplot prepared={
      median=3,
      upper quartile=6,
      lower quartile=2,
      upper whisker=12,
      lower whisker=1,
      every median/.style={darkred,ultra thick},
    },
    ] coordinates {};

  \end{axis}
\end{tikzpicture}
  \caption{Size distribution of interval sets}
  \label{fig:eval-size-dist}
\end{figure}
\subsection{Size Distribution of Interval Set}\label{sec:eval-interval-set-size}
Section~\ref{sec:design-augmented-reachability} introduced the concept of an interval set $\intervalset{v}$ for a vertex $v$ and its subsuming operator $\supseteq$. The properties of an interval set play an important role in efficiently determining the reachability of two vertices. Specifically, given two interval sets $\intervalset{p}$ and $\intervalset{q}$ with at most $N$ intervals each, the time complexity for evaluating $\intervalset{p} \supseteq \intervalset{q}$ is $O(N^2)$.

Figure~\ref{fig:eval-size-dist} shows the size distribution of interval sets across all targets in both client analyses. In taint analysis, half of the interval sets have at most 2 intervals; the maximum value of this chart also tells us that majority of the interval sets have at most 8 intervals. For read-only argument analysis, majority of the interval sets have at most 17 intervals, while half of them have no more than 4 intervals.
These numbers show that the average size of an interval set is usually small. Thus, they offer additional evidence towards the effectiveness of our novel reversed DFT traversal scheme in reducing the number of non-tree edges.

\section{Discussion}
\label{sec:discussion}

\subsection{Fixpoint Convergence}
The intraprocedural reversed DFT traversal, introduced in Section~\ref{sec:design-intraprocedural}, operates basically the same as normal DFS traversal on spanning tree edges and thus ensures its fixpoint convergence on those edges.
While handling non-tree edges, we never remove any interval or interval subset during the propagation, therefore the process is monotonic with respect to the partial orders of the interval sets. Thus, fixpoint convergence is given.
For the interprocedural case, we focus on Algorithm~\ref{algo:defi-interprocedual-worklist}.
The terminating condition for Algorithm~\ref{algo:defi-interprocedual-worklist} is dictated by changes to function summaries. These changes are caused by transitively propagating any callee function summary into the current function context. The propagation is driven by the same traversal algorithm that is also used for the intraprocedural case mentioned above. Thus, a fixpoint convergence w.r.t. function summaries can be inducted from the convergence of interval sets in callee functions, due to the fact that a function summary is derived from interval sets of the same function.

\subsection{Comparison with IFDS}

\coolname{} replaces the IFDS reachability algorithm with a simple interval lookup and avoids the accumulation of PathEdges which constitute a large part of the memory footprint of IFDS solvers~\cite{hePerformanceBoostingSparsificationIFDS2019,arztSustainableSolvingReducing2021,liScalingIFDSAlgorithm2021}.
Further, \coolname{} processes only relevant program statements by utilizing sparse \textit{SSA def-use chains}, whereas IFDS processes every program statement along \textit{control flow paths} inducing memory and runtime overheads~\cite{hePerformanceBoostingSparsificationIFDS2019}.
Lastly, \coolname{} decouples DFT interval computation from specific value-flow queries and can determine value-flows between \textit{arbitrary} nodes.
In contrast, IFDS requires non-trvial number of recomputations on PathEdges~\cite{arztReviserEfficientlyUpdating2014} whenever there is a change in value-flow sources.

%

\section{Related Work}
\label{sec:related}

The original IFDS/IDE algorithm~\cite{reps1995ifds} together with practical extensions proposed by Naeem et al.~\cite{naeemPracticalExtensionsIFDS2010}
is nowadays implemented by many analysis frameworks for JAVA~\cite{boddenInterproceduralDataflowAnalysis2012,arztFlowDroidPreciseContext2014a} and C/C++~\cite{schubertPhASARInterproceduralStatic2019}.
In addition, several approaches have been proposed to further reduce memory consumption and processing time of IFDS implementations.
Sparsedroid~\cite{hePerformanceBoostingSparsificationIFDS2019} improves the sparsity of the dataflow propagation.
The number of dataflow edges is reduced by connecting dataflow facts directly to their next point of use, instead of the next node in the CFG.
DiskDroid~\cite{liScalingIFDSAlgorithm2021} and CleanDroid~\cite{arztSustainableSolvingReducing2021} reduce the footprint of the graph reachability algorithm by detecting stale edges and either move them to disk or completely remove them from the working set.
WALA~\cite{WatsonLibrariesAnalysis} proposes an efficient bitset representation of dataflow facts.
Coyote~\cite{shiPipeliningBottomupData2020} improves the parallelism of bottom-up IFDS implementations by increasing the granularity of caller-callee dependencies.
Intraprocedural analysis is split into multiple independent parts which can then be run in parallel.
All of the mentioned approaches address shortcomings that are specific to the original IFDS implementation and are therefore not directly comparable to \coolname{}.


Recent contributions such as BigSpa~\cite{zuoBigSpaEfficientInterprocedural2019}, Grapple~\cite{zuoGrappleGraphSystem2019}, GraSpan~\cite{wangGraspanSinglemachineDiskbased2017} and Chianina~\cite{zuoChianinaEvolvingGraph2021}
work around scalability issues of static analysis frameworks by developing core functionalities inspired by big data analytics in order to support certain classes of static analyzers.
The actual analysis is then implemented on top of the core API and can thereby be transparently scaled to the available resources of the underlying system.
Systems approaches are orthogonal to \coolname{} as they aim to provide primitives to improve the resource utilization of static analyzers but do not aim to optimize the analysis algorithm itself.

\section{Conclusion}
Value-flow analysis is an important component of many program optimizations. Previous researchers have made significant contributions and created important tools, but true scalability of these tools to large real-world codebases has so far proven elusive.
We present a solution that is able to overcome these scalability bottlenecks. Key to our approach is a dialect of LLVM IR that simplifies the modeling of pointer aliasing for analysis. This leads to a novel sparse intermediate representation that results in graphs with low tree widths. The resulting graph algorithms have much lower resource requirements and much better performance characteristics than previous approaches and provide almost linear scalability to truly large programs. Our prototype implementation is based on LLVM, is source-language agnostic, and will be open-sourced.

\bibliographystyle{ACM-Reference-Format}
\bibliography{refs}

\appendix
\section{Supplements}
\label{sec:appendix}
\subsection{Tables}
\label{sec:appendix:tables}
\begin{table}[h]
    \centering
    \resizebox{\columnwidth}{!}{%
    \begin{tabular}{|l|c|c|c|c|c|}
    \hline
    \multicolumn{1}{|c|}{Benchmark name} & no use & 1 use  & 2 uses & 3 uses & 4+ uses \\ \hline
    libcrypto (OpenSSL)                  & 31.89\% & 59.65\% & 4.00\%  & 1.53\%  & 2.93\%   \\ \hline
    libssl (OpenSSL)                     & 33.44\% & 59.97\% & 2.30\%  & 1.55\%  & 2.74\%   \\ \hline
    SQLite 3                             & 30.70\% & 52.26\% & 9.53\%  & 3.09\%  & 4.42\%   \\ \hline
    FFmpeg                               & 20.67\% & 59.98\% & 11.53\% & 3.59\%  & 4.23\%   \\ \hline
    Lighttpd                             & 30.50\% & 63.74\% & 2.08\%  & 1.05\%  & 2.63\%   \\ \hline
    Servo                                & 37.19\% & 50.58\% & 7.43\%  & 1.69\%  & 3.11\%   \\ \hline
    \end{tabular}%
    }
    \caption{Percentage of SSA values with the respective number of uses}
    \label{tab:ssa-num-value-users}
\end{table}


\subsection{Listings}
\label{sec:appendix:listings}
\begin{lstlisting}[
  style=MLIRStyle,
  captionpos=b, caption={IR before conversion to \texttt{dfi.call}.},
  label={lst:defi-call-before}]
llvm.func @f(%v: i32, %p: !llvm.ptr<i32>) -> i64 {
  %r = llvm.call @g(%v, ?\colorbox{green}{\%p}?)
  %0 = llvm.load ?\colorbox{green}{\%p}? : !llvm.ptr<i32>
  llvm.return %r : i64
}
\end{lstlisting}%
\begin{lstlisting}[
  style=MLIRStyle,
  captionpos=b, caption={IR after conversion to \texttt{dfi.call}. The newly added function call result \texttt{\%p0} on line~2 captures side effects induced by the callee \texttt{g} w.r.t pointer argument \texttt{\%p}.},
  label={lst:defi-call-after}]
llvm.func @f(%v: i32, %p: !llvm.ptr<i32>) -> i64 {
  %r,?\colorbox{yellow}{\%p0}? = dfi.call @g(%v, ?\colorbox{green}{\%p}?)
  %0 = llvm.load ?\colorbox{yellow}{\%p0}? : !llvm.ptr<i32>
  llvm.return %r : i64
}
\end{lstlisting}

\end{document}